\documentclass[12]{article}

\usepackage{amsmath,amssymb,amsthm}
\usepackage{hyperref}
\usepackage{natbib}
\usepackage{mathtools}
\usepackage[title]{appendix}
\usepackage{fullpage}
\usepackage{setspace,rotating}
\usepackage{threeparttable,booktabs}

\newtheorem{lemma}{Lemma}
\newtheorem{theorem}{Theorem}
\newtheorem{corollary}{Corollary}

\newtheorem{proposition}{Proposition}

\newcommand{\sign}{\mathrm{sign}}

\hypersetup{
	colorlinks   = true, 
	urlcolor     = blue, 
	linkcolor    = blue, 
	citecolor   = blue 
}

\begin{document}
	
	\title{Formal Covariate Benchmarking to Bound Omitted Variable Bias}
	
	\author{Deepankar Basu\thanks{Department of Economics, University of Massachusetts Amherst. Email: \texttt{dbasu@econs.umass.edu}. I would like to thank Michael Ash and Carols Cinelli for very helpful comments on an earlier version of the paper. All remaining errors are mine.}}
	
	\date{\today}
	
	\maketitle
	
	\begin{abstract}
		Covariate benchmarking is an important part of sensitivity analysis about omitted variable bias and can be used to bound the strength of the unobserved confounder using information and judgments about observed covariates. It is common to carry out formal covariate benchmarking after residualizing the unobserved confounder on the set of observed covariates. In this paper, I explain the rationale and details of this procedure. I clarify some important details of the process of formal covariate benchmarking and highlight some of the difficulties of interpretation that researchers face in reasoning about the residualized part of unobserved confounders. I explain all the points with several empirical examples.
		
	\end{abstract}
	
	\doublespacing
	
	\section{Introduction}
	In many disciplines, like economics, epidemiology, political science, public health, sociology, etc., it is of utmost importance to estimate causal effects from observational data, e.g. the causal effect of years of schooling on wages \citep{card_2001}, the causal effect of class size on student scores \citep{angrist-lavy-1999}, the causal effect of exposure to violence on attitudes towards peace \citep{cinelli-hazlett-2020}, or the causal effect of breastfeeding on child outcomes \citep{vanderweele-ding-2017}. In each of these cases, and in observational studies more generally, researchers need to take account of unmeasured confounders, i.e. unobserved variables that are correlated both with treatment assignment and the outcome under study, if they wish to distinguish between correlation and causation. When plausible instrumental variables are not available for treatment assignment, as is often the case in observational studies, researchers can turn to sensitivity analysis to investigate the robustness of their results. 
	
	Sensitivity analysis in the context of unobserved confounders, and to deal with omitted variable bias \citep{deluca-2018, basu-2020} resulting therefrom, involves at least two different steps. In the fist step, the researcher tries to quantify the strength of association between the unmeasured confounder and the treatment (and outcome) that could overturn the conclusions of her research. If this magnitude of association is `large', researchers can be relatively confident about the robustness of their results to the presence of unmeasured confounders. While this first step is better than conducting no sensitivity analysis, the following question still needs to be answered: how `large' is too large? Covariate benchmarking, the second step of sensitivity analysis, can help answer this question. 
	
	In the second step of covariate benchmarking, the researcher uses strengths of association between observed covariates and the treatment (and outcome) to reason about the possible magnitudes of association of the unobserved confounder with the treatment and outcome. If, using covariate benchmarking, the researcher can rule out magnitudes of association that was found problematic in the first step, then her case for robustness is significantly strengthened.
	
	There is a large literature on sensitivity analysis that goes back to at least \citet{cornfield-etal-1959}, and has been continued in \citet{rosenbaum-rubin-1983, rosenbaum-2002, imbens-2003, imai-etal-2010}, and others. Among more recent proposal for sensitivity analysis \citep{frank_2000,krauth-2016,ding-vanderweele-2016,vanderweele-ding-2017,oster-2019}, a most innovative and promising approach has been presented by \citet{cinelli-hazlett-2020}. The key novelty in this proposal involved re-parametrizing the traditional omitted variable bias expression using partial $R^2$ measures.\footnote{For a definition of the partial $R^2$, see equation~\ref{def-partialr2} below.} This re-parametrization has opened up a fruitful way to conduct sensitivity analysis about omitted variable bias, a stripped-down version of which involves the following three-step process.
	
	In the first step, the researcher computes \textit{robustness values} of two parameters, which are understood as the magnitude of the minimum strength of association (measured with the partial $R^2$) that an unobserved confounder would need to have, both with the treatment and with the outcome, to change the research conclusions. By themselves, the robustness values are of limited use because the confounder is not observed. Hence, it is difficult for a researcher to judge whether the robustness values are too high or too low for her specific study. In the second step, the researcher addresses this difficulty using \textit{formal covariate benchmarking}---arguable the most difficult and crucial part of the whole analysis.\footnote{Formal covariate benchmarking should be distinguished from informal covariate benchmarking. In the former, formal, quantitative arguments are used; in the latter, informal, non-quantitative arguments are used. \citet[section~4.4]{cinelli-hazlett-2020} argue persuasively that informal covariate benchmarking can often be misleading because it does not solve the correct identification problem.}  In this step, the researcher uses information about an \textit{observed} covariate (or set of covariates) to compute \textit{upper bounds} for the strength of association, measured once again using the partial $R^2$, between the \textit{unobserved} confounder and treatment (and outcome). In the final and third step, the researcher compares the magnitudes of the robustness values computed in the first step with the upper bounds computed in the second step. If the upper bounds are lower than the robustness values, then the researcher is able to conclude that her conclusions are robust to omitted variable bias.
	
	In this elegant methodology, the crucial, second step of formal covariate benchmarking relies on an orthogonality assumption that comes in either of two forms. Either it is assumed that the unobserved confounder is orthogonal to the observed covariates, or it is assumed that researchers can work with the residualized part of the unobserved confounder, i.e. the part of the unobserved confounder that is orthogonal to the set of observed covariates. The main contribution of this paper is to extend or expand the methodology proposed in \citet{cinelli-hazlett-2020} in several ways.
	
	First, I explain why orthogonality assumptions are necessary for formal covariate benchmarking. In particular, in section~\ref{sec:cov-bench-general}, I present a general discussion of sensitivity analysis along the lines of \citet{cinelli-hazlett-2020} and then, in section~\ref{sec:cov-bench-ortho-1}, I explain the role of both forms of orthogonality for formal covariate benchmarking. 
	
	The second contribution of this paper is to clarify an important step of the formal covariate benchmarking methodology: computing the absolute value of the partial correlation of the outcome and the unobserved confounder controlling for the treatment and observed covariates, $|R_{Y \sim Z|D,X}|$. Extending the analysis in \citet[appendix~B]{cinelli-hazlett-2020}, I show in section~\ref{sec:cov-bench-ortho-2} how to determine $|R_{Y \sim Z|D,X}|$ in all cases, not only in the case when ``the confounder acts towards \textit{reducing} the magnitude of the estimate towards zero'' \citet[appendix~B, p.~34,  emphasis in original]{cinelli-hazlett-2020}.
	
	The third contribution of the paper is to highlight that the conclusions of sensitivity analysis differs according to whether researchers use a total or a partial $R^2$-based comparison for covariate benchmarking. This is not surprising because these are two different ways of defining `relative importance' of the unobserved confounder and the observed covariates, and it has been previously pointed out that different concepts of relative importance often affect conclusions of research \citep[p.~2]{kruskal-majors-1989}. But this implies that researchers need to think carefully about which they want to use: total $R^2$-based comparisons, partial $R^2$-based comparison or both. In the examples that I discuss, total $R^2$-based comparisons provide a more conservative approach and this might be used if researchers wish to be cautious about the conclusions of their studies.
	
	The final contribution of this paper is to highlight some of the difficulties of interpretation that researchers face when using the methodology proposed in \citet{cinelli-hazlett-2020}. The difficulties arise from the need to reason about the residualized part of the unobserved confounder. Compounding the difficulty that the confounder is itself unobserved, is the further problem that one has to form reasoned arguments about only a part of this unobserved variable---the part that is not explained linearly by the set of observed covariates. Hence, it is not immediately obvious how one might form reasonable and reliable judgments about the explanatory power of the residualized part of the unobserved confounder. Perhaps this difficulty is unavoidable, but researchers must, at the least, be aware of it.  

	Other recent work on sensitivity analysis that have used the orthogonality assumption are \citet[equation 7, page~121]{krauth-2016} and \citet[page~192]{oster-2019}.\footnote{Also see the exposition of Oster's analysis in \citet{deluca-2019} and note the importance of assumption~B (orthogonality) for theorem~1 \citep[page~219]{deluca-2019}.} I do not discuss these papers because both suffer from problems. In \citet{krauth-2016}, the key sensitivity parameter, $\lambda$, does not have natural bounds. This is because $\lambda$ is the \textit{ratio} of two correlation coefficients, and it is difficult to restrict this ratio to a finite interval on the real line, as \citet[assumption~2]{krauth-2016} does, without additional assumptions. On the other hand, \citet{oster-2019} suffers from the problem that one of the key sensitivity parameters, $\delta$, cannot be interpreted in a way that is useful for sensitivity analysis.\footnote{See \citet[section~6.3]{cinelli-hazlett-2020} for more details.} 
	
	A recent work on sensitivity analysis that does \textit{not} use the orthogonality assumption is \citet{ding-vanderweele-2016} and \citet{vanderweele-ding-2017}. Much like the analysis in \citet{cinelli-hazlett-2020}, these papers propose sensitivity analysis without assuming orthogonality between the unobserved confounder and the observed covariates. But I do not discuss these papers because they do not use formal covariate benchmarking to understand whether the strengths of association that would nullify the estimated effect is reasonable in the context of a specific study. Since the primary goal of my paper is to analyze covariate benchmarking, I do not discuss \citet{ding-vanderweele-2016} and \citet{vanderweele-ding-2017}.
	
	The rest of the paper is organized as follows: in section~\ref{sec:setup}, I present the basic set up and the key expression for bias, and discuss the three-step procedure for sensitivity analysis proposed in \citet{cinelli-hazlett-2020}; in section~\ref{sec:r2-result}, I present some novel results about the total and partial $R^2$ that will be useful later in the paper; in section~\ref{sec:cov-bench-general}, I provide a general discussion of covariate benchmarking; in section~\ref{sec:cov-bench-ortho-1}, I discuss the first step of covariate benchmarking when either form of the orthogonality assumption is used; in section~\ref{sec:cov-bench-ortho-2}, I discuss the second step; in section~\ref{sec:examples}, I discuss several examples of observational studies; I conclude in section~\ref{sec:conclusion}. Proofs are collected in the appendix. Throughout this paper, I will follow the notation used in \citet{cinelli-hazlett-2020} to facilitate easy comparison.

	\section{The Setup}\label{sec:setup}
	
	\subsection{What is partial $R^2$?}
	
	The concept of partial $R^2$ is not very widely used in econometrics.\footnote{I could not find this concept being discussed in any one of the popular graduate-level textbooks on econometrics. The closest one comes is the discussion of the partial correlation coefficient in \citet[section~3.4]{greene}. The classic treatment of partial $R^2$ can be found in \citet[chapter~12]{yule-1911}, or in later editions of the book, e.g. \citet[chapter~14]{yule-kendall-1948}.} Since this concept is central to the analysis in this paper, I begin by discussing it briefly. The partial $R^2$ of the random variables $Y$ and $Z$, conditional on a set of covariates, $X$, can be computed as follows \citep[section~3.4]{greene}: (a) collect the vector of residuals from a regression of $Y$ on $X$; (b) collect the vector of residuals from a regression of $Z$ on $X$; (b) take the square of the correlation coefficient between the two vectors of residuals. This is the partial $R^2$ of the random variables $Y$ and $Z$, conditional on a set of covariates, $X$. While this computation clarifies the \textit{partialling out} involved in defining the partial $R^2$, there are two different, more useful, ways to define it.\footnote{These definitions are given in \citet[page~12]{cinelli-hazlett-2020}.}
	
	The partial $R^2$ between the random variables $Y$ and $Z$, conditional on a set of covariates, $X$, denoted by $R^2_{Y \sim Z|X}$, can also be defined in terms of the more familiar total $R^2$ (the coefficient of determination in a regression), as follows:
	\begin{equation}\label{def-partialr2}
		R^2_{Y \sim Z|X} = \frac{R^2_{Y \sim Z+X} - R^2_{Y \sim X}}{1- R^2_{Y \sim X}},
	\end{equation}
	where $R^2_{Y \sim Z+X}$ denote the total $R^2$ in the regression of $Y$ on $Z$ and $X$, and $R^2_{Y \sim X}$ denotes the total $R^2$ in a regression of $Y$ on $X$. From this definition in (\ref{def-partialr2}) we get some more intuition about what the partial $R^2$ measures: it is the ratio of (a) the increment in the total $R^2$ when $Z$ is added as a covariate to the regression of $Y$ on $X$, and (b) the difference of the total $R^2$ of the regression of $Y$ on $X$ from $1$. Since the total $R^2$ can be at most $1$ and since the total $R^2$ always increases weakly with the addition of a regressor, the numerator can at most be as large as the denominator, with both the numerator and denominators being positive. Hence, it is immediately clear that the partial $R^2$, like the total $R^2$, must lie between $0$ and $1$. 
	
	There is yet another definition of the partial $R^2$ that is motivated by another consideration: to re-express the partial $R^2$ of two random variable by removing one random variable, or several random variables, from the conditioning set. Suppose, for concreteness, that we wish to express the partial $R^2$ of $Y$ and $Z$, conditional on $X$ and $D$, in terms of partial $R^2$ measures conditional only on $X$, i.e. we remove $D$ from the conditioning set. This can be accomplished with the recursive definition of partial $R^2$:
	\begin{equation}\label{def-recursive-partialr2}
		R_{Y \sim Z|D,X} = \frac{ R_{Y \sim Z|X} - R_{Y \sim D|X} R_{D \sim Z|X} }{\sqrt{1-R^2_{Y \sim D|X}} \sqrt{1-R^2_{D \sim Z|X}}}.
	\end{equation}
	I will use (\ref{def-partialr2}) and (\ref{def-recursive-partialr2}) extensively in this paper.
	
	\subsection{Expression of relative omitted variable bias}
	Consider the linear regression of an outcome on a treatment, controlling for a set of covariates given by $X$ and $Z$,
	\begin{equation}\label{model-full}
		Y = \hat{\tau}D + X \hat{\beta} + \hat{\gamma}Z + \hat{\varepsilon}_{\textrm{full}}
	\end{equation}
	where $Y$ is the $n \times 1$ vector of the outcome (dependent variable), $X$ is the $n \times k$ matrix of observed covariates, including a constant, $Z$ is the $n \times 1$ (unobserved) confounder vector, and all hat-quantities denote estimated (sample, and not population) quantities. Since $Z$ is unobserved, the researcher cannot estimate (\ref{model-full}) but is forced to estimate the following restricted regression
	\begin{equation}\label{model-rest}
		Y = \hat{\tau}_{\textrm{res}}D + X \hat{\beta}_{\textrm{res}} + \hat{\varepsilon}_{\textrm{res}}
	\end{equation}
	Letting 
	\begin{equation}\label{def-bias}
		\widehat{\text{bias}} = \hat{\tau}_{\text{res}} - \hat{\tau}
	\end{equation}
	denote the bias of the treatment effect arising from the restricted model, \citet[page~48]{cinelli-hazlett-2020} show, by combining the Frisch-Waugh-Lovell theorem and definitions of partial $R^2$, that
	\begin{equation}\label{bias}
		\left| \widehat{\textrm{bias}}\right| = \textrm{se}\left( \hat{\tau}\right) \sqrt{\frac{\textrm{df} \times R^2_{Y \sim D, X} \times R^2_{D \sim Z|X}}{1-R^2_{D \sim Z|X}}}  
	\end{equation}
	where `se' denotes standard error, `df' denotes the degrees of freedom of the restricted regression in (\ref{model-rest}), $R^2_{Y \sim D, X}$ denotes the total $R^2$ from a regression of $Y$ on $D$ and $X$, $R^2_{D \sim Z|X}$ refers to the partial $R^2$ from a regression of $D$ on $Z$ conditioning on $X$ and we assume that $0 \leq R^2_{D \sim Z|X}<1$ (to make sure we do not attempt to divide by zero). The expression for bias can be further manipulated to derive the expression for `relative bias' \citep[page~49]{cinelli-hazlett-2020}
	\begin{equation}\label{bias-rel}
		\left| \frac{\widehat{\textrm{bias}}}{\hat{\tau}_{\textrm{res}}}\right| = \frac{\left| R_{Y \sim Z|D,X} \right| \times \left| R_{D \sim Z|X}\right|}{\left| R_{Y \sim D|X}\right| } \sqrt{ \frac{1-R^2_{Y \sim D|X}}{1-R^2_{D \sim Z|X}} },
	\end{equation}
	which will be of use in sensitivity analysis. \footnote{In this paper, I will only deal with the case of a single confounder. The case of multiple confounders does not need a separate treatment because the bias with a single confounder is an upper bound for the bias with multiple confounders \citep[section~4.5]{cinelli-hazlett-2020}.}
	
	\subsection{Importance of relative bias for sensitivity analysis}
	What is the importance of relative bias for sensitivity analysis? Relative bias is crucial because it helps a researcher address the question whether taking account of omitted variable bias can overturn the conclusions of an observational study. For, if relative bias is equal to or larger than $1$, then the magnitude of the bias can large enough to nullify any nonzero treatment effect that might have been estimated by a researcher, i.e. she would not be able to rule out the possibility that taking account of omitted variable bias would make the estimated treatment effect zero. Sensitivity analysis will, therefore, boil down to seeing if relative bias is larger or smaller than unity. Hence, to investigate how sensitive an estimate of a treatment effect is to omitted variable bias, a researcher should analyze the conditions under which relative bias might equal or exceed unity. Drawing on \citet{cinelli-hazlett-2020}, we might propose a three-step sensitivity analysis for this purpose.  
	
	\subsection{Sensitivity analysis}
	
	\subsubsection{Step~1: Compute robustness values}\label{sec:step1}
	The first step of the sensitivity analysis is to compute two `robustness values'. The first robustness value is $RV_q$, which asks us to answer the following question: If the partial $R^2$ of the confounder with the treatment, $R^2_{D \sim Z|X}$ and of the confounder with the outcome $R^2_{Y \sim Z|D,X}$ were equal in magnitude, how strong would this partial $R^2$ need to be to reduce the estimated treatment effect by $100\times q\%$. If $q=1$, this open up the possibility that the treatment effect is zero, the case most commonly of interest to researchers. In the rest of this paper, I will focus on this case and use $RV$ to denote $RV_1$.
	
	The second robustness value is $RV_{q, \alpha}$, which answers the following question: If the partial $R^2$ of the confounder with the treatment, $R^2_{D \sim Z|X}$ and of the confounder with the outcome $R^2_{Y \sim Z|D,X}$ were equal in magnitude, how strong would this partial $R^2$ need to be to make the adjusted t-test \textit{not reject} the null hypothesis that the true treatment effect is $(1-q)|\hat{\tau}|$ at the $\alpha$ level? Thus, $RV_{q, \alpha}$ allows a researcher to see whether the estimated treatment effect is zero, after taking account of the uncertainty associated with estimation. My focus will be on the case with $q=1$, where $RV_{q, \alpha}$ will refer to the null hypothesis that the true treatment effect is zero. For notational simplicity, $RV_{1,\alpha}$ is denoted simply as $RV_{\alpha}$.
	
	\subsubsection{Step~2: Compute bounds using covariate benchmarking}
	The second step of the sensitivity analysis is the difficult and crucial step. 
	\begin{quote}
		Arguably, the most difficult part of a sensitivity analysis is taking the description of a confounder that would be problematic from the formal results [e.g., the robustness values], and reasoning about whether a confounder with such strength plausibly exists in one’s study, given its design and the investigator’s contextual knowledge. \citep[page~13]{cinelli-hazlett-2020}.
	\end{quote}
	In this step, the researcher needs to investigate the question whether she can reasonably rule out the possibility that $R^2_{D \sim Z|X}$ and $R^2_{Y \sim Z|D,X}$ are higher than the robustness values. Since these two partial $R^2$ values cannot be computed---because $Z$ is unobserved---she must use information about \textit{observed} covariates to find exact expressions or upper bounds for them. This is where formal covariate benchmarking comes in.
	
	At this point, \citet{cinelli-hazlett-2020} introduce two parameters, $k_D$ and $k_Y$, to assist in the process. The first parameter, $k_D$, captures the relative strength of the confounder in explaining variation in the treatment as compared to a chosen, observed covariate (or set of covariates); the second parameter, $k_Y$, captures the corresponding relative strength of the unobserved confounder for explaining variation in the outcome. Both parameters can be defined with and without conditioning on observed covariates and the treatment. These parameters capture the judgment of the researcher based on her knowledge of the context of the research. Once the values of $k_D$ and $k_Y$ have been chosen, \citet{cinelli-hazlett-2020} show that we can generate exact expressions for $R^2_{D \sim Z|X}$ and $R^2_{Y \sim Z|D,X}$ as functions of known quantities and $k_D$ (or $k_Y$).  
	
	\subsubsection{Step~3: Compare robustness values with bounds}\label{sec:step3} 
	In the third and final step, the researcher needs to compare the magnitudes of $R^2_{D \sim Z|X}$ and $R^2_{Y \sim Z|D,X}$, computed in the second step, with the magnitudes of the robustness values computed in the first step. This comparison can then allow the researcher to assess the robustness of the results to omitted variable bias. In particular:
	\begin{enumerate}
		\item If 
		\[
		\max \left\lbrace R^2_{D \sim Z|X},R^2_{Y \sim Z|D,X}\right\rbrace < RV
		\]
		then the researcher can conclude that the point estimate of the treatment effect is robust to omitted variable bias;
		
		\item If 
		\[
		\max \left\lbrace R^2_{D \sim Z|X},R^2_{Y \sim Z|D,X}\right\rbrace < RV_{\alpha}
		\]
		then the researcher can conclude that the bias-adjusted t-test of the null hypothesis that the treatment effect is zero can be rejected at the $\alpha\%$ level of significance;
		
		\item If 
		\[
		R^2_{D \sim Z|X} < R^2_{Y \sim D|X}
		\]
		then the researcher can conclude that the ``worst case confounder'' (a confounder that explains all the residual variance in the outcome, i.e. $R^2_{Y \sim Z|D,X}=1$) would not eliminate the the estimated treatment effect.
	\end{enumerate}

	\section{A result about total and partial $ R^2 $ }\label{sec:r2-result}
	For any $n \times k$ matrix, $W$, let	
	\[
	P_W = W\left(W'W \right)^{-1}W', 
	\]    
	denote the projection matrix that projects onto the column space of $W$; let $M^0$ denote the $n \times n$ matrix that generates deviations from means when pre-multiplied to a $n$ vector \citep[page~978--79]{greene}, i.e.,
	\[
	M^0 = \left[ I - \frac{1}{n} i i'\right],
	\]
	where $I$ is the identity matrix of dimension $n$ and $i$ denotes a column vector of $1$s. Note that $P_W$ and $M^0$ are symmetric and idempotent matrices \citep[page~32, 979]{greene}.

	For a $n \times 1$ vector, $Z$, let $Z^{\perp X} = Z - P_XZ = \left( I-P_X\right)Z $ denote the $n \times 1$ vector of ordinary least squares (OLS) residuals obtained from a regression of $Z$ on $X$, and consider the following four regressions estimated by OLS:
	\begin{align}
		Y & \textrm{ on } X, Z \label{reg:yxz}\\
		Y & \textrm{ on } X \label{reg:yx}\\
		Y & \textrm{ on } Z \label{reg:yz}\\
		Y & \textrm{ on } Z^{\perp X} \label{reg:yz^x}
	\end{align}
	Let $R^2_{Y \sim X+Z}$, $R^2_{Y \sim X}$, $R^2_{Y \sim Z}$, and $R^2_{Y \sim Z^{\perp X}}$, denote the total R-squared (coefficient of determination) for the regressions in (\ref{reg:yxz}), (\ref{reg:yx}), (\ref{reg:yz}), and (\ref{reg:yz^x}), respectively; and let $W = \left( X:Z\right) $ denote the $n \times (k+1)$ augmented matrix obtained by appending $Z$ as an additional column to the matrix $X$. Using the definition of the R-squared \citep[page~41]{greene}, we have
	\begin{align}
		R^2_{Y \sim X+Z} & = \frac{\left( P_W Y\right)' M^0\left( P_W Y\right) }{Y'M^0 Y} = \frac{Y' P_W M^0 P_W Y}{Y' M^0 Y} \label{r2:yxz} \\
		R^2_{Y \sim X} & = \frac{\left( P_X Y\right)' M^0 \left( P_X Y\right) }{Y' M^0 Y} = \frac{Y' P_X M^0 P_X Y}{Y' M^0 Y}\label{r2:yx} \\
		R^2_{Y \sim Z} & = \frac{\left( P_Z Y\right)' M^0 \left( P_Z Y\right) }{Y' M^0 Y} = \frac{Y' P_Z M^0 P_Z Y}{Y' M^0 Y}\label{r2:yz} \\
		R^2_{Y \sim Z^{\perp X}} & = \frac{\left( P_Z^{\perp X} Y\right)' M^0 \left( P_Z^{\perp X} Y\right) }{Y' M^0 Y} = \frac{Y' P_Z^{\perp X} M^0 P_Z^{\perp X} Y}{Y' M^0 Y}\label{r2:yz^x} 
	\end{align}
	where $P_W, P_X, P_Z$, and $P_{Z^{\perp X}}$ denote $n \times n$ projection matrices onto the column spaces of $W, X, Z$, and $Z^{\perp X}$ respectively. I will need a result on the decomposition of projection matrices that is given in \citet[page~323]{rao-etal-2008}.
		\begin{lemma}\label{lemma:decomp}
			The projection matrix of $W$ can be decomposed into two projection matrices as:
			\begin{equation}\label{eq:proj-decomp}
				P_W = P_X + P_{Z^{\perp X}}. 
			\end{equation}
		\end{lemma}
Using lemma~\ref{lemma:decomp}, we can prove the following result about the decomposition of the total $R^2$. 		
	\begin{theorem}\label{thm:r2-decomp}
	For the regressions in (\ref{reg:yxz}), (\ref{reg:yx}), (\ref{reg:yz}), and (\ref{reg:yz^x}), we have:
	\begin{equation}\label{r2decomp-1}
		R^2_{Y \sim X+Z} = R^2_{Y \sim X} + R^2_{Y \sim Z^{\perp X}} = R^2_{Y \sim X+Z^{\perp X}},
	\end{equation}
	and
	\begin{equation}\label{r2decomp-2}
		R^2_{Y \sim X+Z} - R^2_{Y \sim X} - R^2_{Y \sim Z} = \eta_{Y,Z,X},
	\end{equation}					
	where 	
	\begin{equation}\label{def:etaxyz}
		\eta_{Y,Z,X} = R^2_{Y \sim Z^{\perp X}} - R^2_{Y \sim Z}.
	\end{equation}
	\end{theorem}
	
Theorem~\ref{thm:r2-decomp} shows that the total $R^2$ from a regression of $Y$ on $X$ and $Z$ can be decomposed in two ways. In (\ref{r2decomp-1}), it is decomposed into the total $R^2$ from a regression of $Y$ on $X$, and the total $R^2$ from a regression of $Y$ on $Z^{\perp X}$ (the part of $Z$ that is orthogonal to $X$). In (\ref{r2decomp-2}), it is decomposed into three terms: the total $R^2$ from a regression of $Y$ on $X$, the total $R^2$ from a regression of $Y$ and $Z$, and a remainder, $\eta_{Y,Z,X}$. 

\begin{corollary}\label{r2decomp-ortho}
	If $Z \perp X$, then $R^2_{Y \sim X+Z} = R^2_{Y \sim X} + R^2_{Y \sim Z}$.
\end{corollary}
\begin{proof}
	If $Z \perp X$, then $Z =Z^{\perp X}$. Using this in (\ref{r2decomp-1}), we get the desired result. 
\end{proof}
\begin{corollary}\label{cor-partial-r2}
	For the regressions in (\ref{reg:yxz}), (\ref{reg:yx}), and (\ref{reg:yz^x}), we have
	\begin{equation}
		R^2_{Y \sim Z|X} = \frac{R^2_{Y \sim X+Z} - R^2_{Y \sim X}}{1- R^2_{Y \sim X}} = \frac{R^2_{Y \sim X+Z^{\perp X}} - R^2_{Y \sim X}}{1- R^2_{Y \sim X}} = R^2_{Y \sim Z^{\perp X}|X}\label{r2zequal}
	\end{equation}
\end{corollary}
\begin{proof}
	From (\ref{r2decomp-1}), we have $R^2_{Y \sim X+Z} = R^2_{Y \sim X} + R^2_{Y \sim Z^{\perp X}}$. The right hand side is equal to the total $R^2$ from a regression of $Y$ on $X$ and $Z^{\perp X}$ because $Z^{\perp X}$ is orthogonal to $X$ by construction. Hence, $R^2_{Y \sim X+Z} = R^2_{Y \sim X+Z^{\perp X}}$. Using the definition of partial $R^2$ of $Y$ and $Z$ conditional on $X$, we have
	\[
	R^2_{Y \sim Z|X} = \frac{R^2_{Y \sim X+Z} - R^2_{Y \sim X}}{1- R^2_{Y \sim X}} 
	\]
	Now replacing $R^2_{Y \sim X+Z}$ with $ R^2_{Y \sim X+Z^{\perp X}} $, we have the desired result.
\end{proof}

The two corollaries will be useful for the subsequent analysis. Corollary~\ref{r2decomp-ortho} will be useful for sensitivity analysis under the assumption that $Z \perp X$. Corollary~\ref{cor-partial-r2} will be useful for sensitivity analysis when we use the residualization of $Z$ on $X$.

	\section{Covariate benchmarking: A general treatment}\label{sec:cov-bench-general}
	The decision rules laid out in section~\ref{sec:step3} that will allow us to assess the robustness of results to omitted variable bias involves two partial $R^2$ that depend on the unobserved confounder, $Z$: $R^2_{D \sim Z|X}$ and $R^2_{Y \sim Z|D,X}$. While covariate benchmarking will allow us to determine $R^2_{D \sim Z|X}$ and $R^2_{Y \sim Z|X}$ in terms of $k_D$, $k_Y$ and other known quantities, a further step will be needed to find $R^2_{Y \sim Z|D,X}$. Using the recursive definition of partial $R^2$ in (\ref{def-recursive-partialr2}), we have
	\begin{equation}\label{partialr2-rec}
	\left| R_{Y \sim Z|D,X} \right| = \frac{\left| R_{Y \sim Z|X} - R_{Y \sim D|X} R_{D \sim Z|X} \right| }{\sqrt{1-R^2_{Y \sim D|X}} \sqrt{1-R^2_{D \sim Z|X}}}.
	\end{equation}
	While $R_{Y \sim D|X}$ is known, and covariate benchmarking will allow us to determine $| R_{D \sim Z|X}| $ and $| R_{Y \sim Z|X}| $, we will still have to determine $| R_{Y \sim Z|X} - R_{Y \sim D|X} R_{D \sim Z|X} |$. Our strategy will be to express $| R_{Y \sim Z|X} - R_{Y \sim D|X} R_{D \sim Z|X} |$ in terms of $| R_{Y \sim D|X}| $, $| R_{D \sim Z|X}| $ and $| R_{Y \sim Z|X}| $. This is a crucial step of the argument that we will return to below in section~\ref{sec:cov-bench-ortho-2} in greater detail. But first let us see how we might go about finding exact expressions or upper bounds for $R^2_{D \sim Z|X}$ and $R^2_{Y \sim Z|X}$.

	\subsection{Total $ R^2 $-based covariate benchmarking}\label{sec:r2total}
	Suppose there are $j$ observed covariates, $\left\lbrace X_1, X_2, \ldots, X_j\right\rbrace $, and the researcher wishes to use the $j$-th one, $X_j$, for benchmarking. Following  \citet[equation 50, appendix~B.1]{cinelli-hazlett-2020}, let us define two parameters that are based on comparisons of total $R^2$ measures:
	\begin{equation}\label{defkd}
		k_D \coloneqq \frac{R^2_{D \sim Z}}{R^2_{D \sim X_j}}; \qquad k_Y \coloneqq \frac{R^2_{Y \sim Z}}{R^2_{Y \sim X_j}}
	\end{equation}
	Here, $k_D$ is the ratio of the total $R^2$ in a regression of $D$ on $Z$ and the the total $R^2$ in a regression of $D$ on $X_j$. Thus, $k_D$ captures the relative importance of the unobserved confounder in explaining variation in treatment assignment (the numerator), compared to the observed covariate, $X_j$ (the denominator), where relative importance is judged in terms of the total $R^2$, and we assume that $ R^2_{D \sim X_j} >0$. Thee second parameter, $k_Y$, is defined in a similar manner, and captures the relative importance of the unobserved confounder, compared to the observed covariate, $X_j$, in explaining variation in the outcome. 
	
	Can we find an exact expression, or an upper bound, for $ R^2_{D \sim Z|X} $? Using the definition of partial $R^2$ \citep[equation 17, page~51]{cinelli-hazlett-2020}, the result in (\ref{r2decomp-1}), and the definition of $k_D$ in (\ref{defkd}), we get,
	\begin{equation}
		R^2_{D \sim Z|X} = \frac{R^2_{D \sim Z+X} - R^2_{D \sim X}}{1- R^2_{D \sim X}} = \frac{R^2_{D \sim Z} + \eta_{D,Z,X}}{1- R^2_{D \sim X}}= \frac{k_D R^2_{D \sim X_j}}{1-R^2_{D \sim X}} + \frac{\eta_{D,Z,X}}{1-R^2_{D \sim X}} \label{r2dzxbound}
	\end{equation}
	Since $\eta_{D,Z,X} = R^2_{Y \sim Z^{\perp X}} - R^2_{Y \sim Z}$, it is the difference of two positive quantities. Hence, in general, the sign of $\eta_{D,Z,X}$ is indeterminate. This implies that the exact magnitude of $ R^2_{D \sim Z|X} $ in (\ref{r2dzxbound}) is indeterminate too. Thus, we cannot find a useful upper bound for $ R^2_{D \sim Z|X} $, without more information. A similar argument shows that we cannot find a useful upper bound for $ R^2_{Y \sim Z|X} $, without more information.

	\subsection{Partial $R^2$-based covariate benchmarking}\label{sec:r2-partial}
	The partial $R^2$-based formal covariate benchmarking comes in two versions. In the first version, bounding of $R_{Y \sim Z|D,X}$ use comparisons that do \textit{not} condition on $D$ (the treatment variable); in the second version, bounding of $R_{Y \sim Z|D,X}$ use comparisons that condition on $D$ (the treatment variable). I will limit my analysis to the first version and note that it can be easily expanded to cover the second version too.
	
	\subsubsection{Single covariate used for benchmarking}
	We would like, as in the total $R^2$ case, to generate upper bounds for $R^2_{D \sim Z|X}$ and $R^2_{Y \sim Z|X}$. Suppose, like before, there are $j$ covariates, $\left\lbrace X_1, X_2, \ldots, X_j\right\rbrace $, and the researcher wishes to use the $j$-th observed covariate, $X_j$, for benchmarking. Let $X_{-j}$ refer to the set of observed covariates that is \textit{not} used for benchmarking, and following \citet[equation 59, appendix~B.2]{cinelli-hazlett-2020}, let us define two parameters based on comparisons of partial $R^2$ measures:
	\begin{equation}\label{kd-partial-single}
		k_D \coloneqq \frac{R^2_{D \sim Z|X_{-j}}}{R^2_{D \sim X_j|X_{-j}}}; \qquad k_Y \coloneqq \frac{R^2_{Y \sim Z|X_{-j}}}{R^2_{Y \sim X_j|X_{-j}}}
	\end{equation}
	Here, $k_D$ captures the relative importance of the unobserved confounder in explaining variation in treatment assignment, compared to the observed covariate, $X_j$, conditional on $X_{-j}$, where, relative importance is now judged using the partial $R^2$ measures, and we assume that $R^2_{D \sim X_j|X_{-j}}>0$; $k_Y$ can be interpreted similarly to capture the relative importance of the unobserved confounder, compared to the observed covariate, $X_j$, to explain variation in the outcome, conditional on $X_{-j}$.
	
	\begin{theorem}\label{thm:r2dzx}
		Suppose $k_D$ and $k_Y$ are defined in (\ref{kd-partial-single}), $0 \leq R^2_{Z \sim X_j | X_{-j}}<1$, $0 \leq R^2_{D \sim X_j | X_{-j}}<1$ and $0 \leq R^2_{Y \sim X_j | X_{-j}}<1$. Then, we have the following lower bounds for $R^2_{D \sim Z|X}$ and $R^2_{Y \sim Z|X}$, respectively:	
		\begin{equation}\label{r2dzx-lbd}
			R^2_{D \sim Z|X} \geq \alpha_D f^2_{D \sim X_j|X_{-j}}, \quad R^2_{Y \sim Z|X} \geq \alpha_Y f^2_{D \sim X_j|X_{-j}},
		\end{equation}
		where
		\[
		\alpha_D = \frac{\left( \sqrt{k_D} - \left|  R_{Z \sim X_j | X_{-j}}\right|  \right)^2 }{1-R^2_{Z \sim X_j | X_{-j}}} \geq 0; \quad \alpha_Y = \frac{\left( \sqrt{k_Y} - \left|  R_{Z \sim X_j | X_{-j}}\right|  \right)^2 }{1-R^2_{Z \sim X_j | X_{-j}}} \geq 0.
		\]

	\end{theorem}		
	The implication of theorem~\ref{thm:r2dzx} is particularly damaging for sensitivity analysis because here we have lower bounds for $R^2_{D \sim Z|X}$ and $R^2_{Y \sim Z|X}$. For the sensitivity analysis we instead need upper bounds. Theorem~\ref{thm:r2dzx} shows that we cannot generate an upper bound for $R^2_{D \sim Z|X}$ and $R^2_{Y \sim Z|X}$, without more information. 
	
	\subsubsection{Multiple covariates used for benchmarking}\label{r2-multiple}
	If a researcher wishes to use multiple observed covariates for benchmarking, then she will have to face, in addition to the problem of not being able to generate an upper bound, an infeasible bounding exercise. This is because repeated use of the recursive definition of partial $R^2$ will become increasingly complicated and unwieldy. I will demonstrate the argument for $R^2_{D \sim Z|X}$ and note that it will equally well apply to $R^2_{Y \sim Z|X}$.
	
	Let $X_{(1,2, \ldots, j)}$ denote the set of covariates that will be used for benchmarking; and let $X_{-(1,2, \ldots, j)}$ denote the complement of that set. Following \citet[equation 66, appendix~B.2]{cinelli-hazlett-2020}, we can define
	\begin{equation}\label{kd-partial-multiple}
			k_D \coloneqq \frac{R^2_{D \sim Z|X_{-(1,2, \ldots, j)}}}{R^2_{D \sim X_{(1,2 \ldots, j)}|X_{-(1,2, \ldots, j)}}}; \qquad k_Y \coloneqq \frac{R^2_{Y \sim Z|X_{-(1,2, \ldots, j)}}}{R^2_{Y \sim X_{(1,2 \ldots, j)}|X_{-(1,2, \ldots, j)}}}
	\end{equation}
	as the parameters to capture the relative strength of the unobserved confounder compared to the set of covariates, $X_{(1,2, \ldots, j)}$, for the bounding exercise. To generate the required bound, we need to apply the recursive definition of partial $R^2$ to $R_{D \sim Z|X}$, $R_{D \sim Z|X_{-(1)}}$, $R_{D \sim Z|X_{-(1,2)}}$, up to $R_{D \sim Z|X_{-(1,2, \ldots , j)}}$.
	
	The first step of the recursion is
	\begin{equation}\label{recursion1-1}
		R_{D \sim Z|X} = \frac{R_{D \sim Z|X_{-(1)}} - R_{D \sim X_{(1)}|X_{-(1)}} R_{Z \sim X_{(1)}|X_{-(1)}} }{\sqrt{1-R^2_{D \sim X_{(1)}|X_{-(1)}}} \sqrt{1-R^2_{Z \sim X_{(1)}|X_{-(1)}}}}
	\end{equation}	

	We can now move to the next step of the recursive process by applying the recursive formula for partial $R^2$ to the numerator of (\ref{recursion1-1}):	
	\begin{equation}\label{recursion2-1}
		R_{D \sim Z|X_{-(1)}} = \frac{R_{D \sim Z|X_{-(1,2)}} - R_{D \sim X_{(2)}|X_{-(1,2)}} R_{Z \sim X_{(2)}|X_{-(1,2)}} }{\sqrt{1-R^2_{D \sim X_{(2)}|X_{-(1,2)}}} \sqrt{1-R^2_{Z \sim X_{(2)}|X_{-(1,2)}}}}
	\end{equation}	

	Plugging (\ref{recursion2-1}) into (\ref{recursion1-1}), we get
	\begin{align*}
		R_{D \sim Z|X} & = \frac{R_{D \sim Z|X_{-(1,2)}}}{\sqrt{1-R^2_{D \sim X_{(2)}|X_{-(1)}}} \sqrt{1-R^2_{Z \sim X_{(2)}|X_{-(1,2)}}} \sqrt{1-R^2_{D \sim X_{(1)}|X_{-(1)}}} \sqrt{1-R^2_{Z \sim X_{(1)}|X_{-(1)}}}} \\
		& \quad - \frac{R_{D \sim X_{(2)}|X_{-(1,2)}} R_{Z \sim X_{(2)}|X_{-(1,2)}} }{\sqrt{1-R^2_{D \sim X_{(2)}|X_{-(1)}}} \sqrt{1-R^2_{Z \sim X_{(2)}|X_{-(1,2)}}} \sqrt{1-R^2_{D \sim X_{(1)}|X_{-(1)}}} \sqrt{1-R^2_{Z \sim X_{(1)}|X_{-(1)}}}} \\
		& \quad - \frac{R_{D \sim X_{(1)}|X_{-(1)}} R_{Z \sim X_{(1)}|X_{-(1)}} }{\sqrt{1-R^2_{D \sim X_{(1)}|X_{-(1)}}} \sqrt{1-R^2_{Z \sim X_{(1)}|X_{-(1)}}}}
	\end{align*}
	If one wished to use covariate benchmarking to replace the terms involving $Z$ in the above expression, one would run into the problem of a proliferation of parameters. Moreover, carrying out the analysis to the next step of the recursion quickly becomes unwieldy. Thus, without more information, the benchmarking exercise using multiple covariates becomes infeasible.
	
	\subsection{Summary}
	In this section, we have seen that covariate benchmarking cannot generate exact expressions or upper bounds for $R_{D \sim Z|X}$ and $R_{Y \sim Z|X}$ unless more information is used about the relationship between $X$ and $Z$. Without these expressions, we cannot also get an exact expression or generate an bound for $R_{Y \sim Z|D,X}$, and sensitivity analysis cannot be conducted. To proceed, we must impose \textit{some} assumption about the relationship between $X$ and $Z$. 
	
	One specific form of this relationship that has proved useful is orthogonality, and I explore it in the next two sections in the two forms that has been used in \citet[page~14]{cinelli-hazlett-2020}: either assume that the unobserved confounder is orthogonal to the set of observed covariates, $Z \perp X$, or work with the residualized part of the unobserved confounder, $Z^{\perp X}$. The assumption that the unobserved confounder is orthogonal to the set of observed covariates is, of course, unrealistic. It is only used because it facilitates derivation of results that can then be used for analyzing the more realistic and useful case where researchers reason about the residualized part of the unobserved confounder. The latter case imposes no formal restrictions because we an always residualize the unobserved confounder with respect to the set of observed covariates.
	
	In terms of the argument to generate relevant bounds, I will, in a first step, derive exact expressions for $R^2_{D \sim Z|X}$ and $R^2_{Y \sim Z|X}$; in the second step, I will derive an exact expression for $R^2_{D \sim Z|D,X}$.
			
	\section{Covariate benchmarking with orthogonality: First step}\label{sec:cov-bench-ortho-1}

	\subsection{$Z$ is orthogonal to $X$}\label{sec:ortho-1}
	
	\subsubsection{Total $R^2$ benchmarking}
	If $Z$ is orthogonal to $X$, then $Z^{\perp X} = Z$. Hence, from (\ref{def:etaxyz}) we see that $\eta_{D,X,Z}=R^2_{D \sim Z^{\perp X}}-R^2_{D \sim Z}=0$. Hence, from (\ref{r2dzxbound}), we see that 
	\[
	R^2_{D \sim Z|X} =  k_D \left( \frac{R^2_{D \sim X_j}}{1-R^2_{D \sim X}}\right), 
	\]
	which gives us an expression for $R^2_{D \sim Z|X}$, with $k_D$ defined in (\ref{defkd}). A similar argument generates an analogous expression for $R^2_{y \sim Z|X}$ in terms of $k_Y$, 
	\[
	R^2_{Y \sim Z|X} =  k_Y \left( \frac{R^2_{Y \sim X_j}}{1-R^2_{Y \sim X}}\right), 
	\]
	with $k_Y$ also defined in (\ref{defkd}).\footnote{This case has been analyzed in \citet[appendix~B.1, page 33-34]{cinelli-hazlett-2020}.} 
	
	\subsubsection{Partial $R^2$ benchmarking using one covariate}
	If $Z$ is orthogonal to $X$, then $ R_{Z \sim X_j|X_{-j}} =0$, and we no longer have an inequality in the proof of theorem~\ref{thm:r2dzx}. Instead, working through the steps, we get
	\[
	R^2_{D \sim Z|X} = k_D \left( \frac{R^2_{D \sim X_j|X_{-j}}}{1-R^2_{D \sim X_j|X_{-j}}}\right),
	\]
	which gives us the required upper bound for $R^2_{D \sim Z|X}$, with $k_D$ defined in (\ref{kd-partial-single}). A similar argument gives us an analogous expression for $R^2_{Y \sim Z|X}$ in terms of $k_Y$, 
	\[
	R^2_{Y \sim Z|X} = k_Y \left( \frac{R^2_{Y \sim X_j|X_{-j}}}{1-R^2_{Y \sim X_j|X_{-j}}}\right),
	\]
	with $k_Y$ defined in (\ref{kd-partial-single}).\footnote{I have omitted the details because this case has been analyzed in \citet[appendix~B.1, page 34-35]{cinelli-hazlett-2020}.} 
	
	\subsubsection{Partial $R^2$ benchmarking using multiple covariates}
	If $Z$ is orthogonal to $X$, then $R^2_{Z \sim X_{(1)}|X_{-(1)}}=0$. Hence, the second term on the numerator and denominator of (\ref{recursion1-1}) are zero, and we get
	\begin{equation}\label{recursion1-2}
			R_{D \sim Z|X} = \frac{R_{D \sim Z|X_{-(1)}}}{\sqrt{1-R^2_{D \sim X_{(1)}|X_{-(1)}}}}
	\end{equation}
	We can now move to the next step of the recursive process by applying the recursive formula for partial $R^2$ to the numerator of (\ref{recursion1-2}):	
	\begin{equation}\label{recursion2-1}
			R_{D \sim Z|X_{-(1)}} = \frac{R_{D \sim Z|X_{-(1,2)}} - R_{D \sim X_{(2)}|X_{-(1,2)}} R_{Z \sim X_{(2)}|X_{-(1,2)}} }{\sqrt{1-R^2_{D \sim X_{(2)}|X_{-(1,2)}}} \sqrt{1-R^2_{Z \sim X_{(2)}|X_{-(1,2)}}}}
	\end{equation}	
	If $Z \perp X$, then $R^2_{Z \sim X_{(2)}|X_{-(1,2)}}=0$. Hence, the second term on the numerator and denominator is zero, once again, and we get
	\begin{equation}\label{recursion2-2}
			R_{D \sim Z|X_{-(1)}} = \frac{R_{D \sim Z|X_{-(1,2)}}}{\sqrt{1-R^2_{D \sim X_{(2)}|X_{-(1,2)}}}}
	\end{equation}
	Plugging (\ref{recursion2-2}) into (\ref{recursion1-2}), we get
	\begin{equation}
			R_{D \sim Z|X} = \frac{R_{D \sim Z|X_{-(1,2)}}}{\sqrt{1-R^2_{D \sim X_{(1)}|X_{-(1)}}} \sqrt{1-R^2_{D \sim X_{(2)}|X_{-(1,2)}}}}
	\end{equation}
	Repeating this process $j$ times, we get 
	\begin{equation}
			R_{D \sim Z|X} = \frac{R_{D \sim Z|X_{-(1,2, \ldots, j)}}}{\sqrt{1-R^2_{D \sim X_{(1)}|X_{-(1)}}} \sqrt{1-R^2_{D \sim X_{(2)}|X_{-(1,2)}}} \cdots \sqrt{1-R^2_{D \sim X_{(j)}|X_{-(1,2, \ldots, j)}}}}
	\end{equation}
	which is equation (67) in \citet[appendix]{cinelli-hazlett-2020}. Noting that the denominator can be simplified to $\sqrt{1-R^2_{D \sim X_{(1,2, \ldots j)}|X_{-(1,2, \ldots, j)}}}$, we get the required upper bound for $R_{D \sim Z|X}$ as
	\[
	R^2_{D \sim Z|X} = k_D \left( \frac{R^2_{D \sim X_{(1,2, \ldots j)}|X_{-(1,2, \ldots, j)}}}{1-R^2_{D \sim X_{(1,2, \ldots j)}|X_{-(1,2, \ldots, j)}}}\right)
	\]
	with $k_D$ defined in (\ref{kd-partial-multiple}). 
	
	A similar argument gives us a similar expression for $R^2_{Y \sim Z|X}$ involving $k_Y$, 
	\[
	R^2_{Y \sim Z|X} = k_Y \left( \frac{R^2_{Y \sim X_{(1,2, \ldots j)}|X_{-(1,2, \ldots, j)}}}{1-R^2_{Y \sim X_{(1,2, \ldots j)}|X_{-(1,2, \ldots, j)}}}\right),
	\]
	with $k_Y$ defines in (\ref{kd-partial-multiple}).\footnote{This case was analyzed in \citet[appendix~B.1, page 36]{cinelli-hazlett-2020}.} 
	
	\subsection{$Z$ is residualized with respect to $X$}\label{sec:ortho-2}
	When the unobserved confounder, $Z$, is residualized with respect to the set of observed covariates, $X$, researchers work with $Z^{\perp X}$, instead of $Z$. Recall that $Z^{\perp X}=Z-P_X Z$ is the vector of residuals from a regression of $Z$ on $X$ estimated by ordinary least squares. Hence, researchers only reason about the part of the unobserved confounder that is orthogonal to $X$, i.e. $Z^{\perp X}$, and not about the unobserved confounder, $Z$, as such. In formal terms, $Z^{\perp X}$ is orthogonal to $X$ by construction. Hence, all the results that were derived under the assumption of orthogonality between $Z$ and $X$ hold once the parameters $k_D$ and $k_Y$ are \textit{suitably redefined and reinterpreted}. 
	
	\subsubsection{Total $R^2$ benchmarking}\label{sec:totalr2-resid} 
	
	Note, first of all, that the definition of $k_D$ and $k_Y$ now uses $Z^{\perp X}$, instead of $Z$:
	\begin{equation}\label{defkd-zperpx}
		k_D \coloneqq \frac{R^2_{D \sim Z^{\perp X}}}{R^2_{D \sim X_j}}, \quad k_Y \coloneqq \frac{R^2_{Y \sim Z^{\perp X}}}{R^2_{D \sim X_j}}.
	\end{equation}
	The interpretation of $k_D$ and $k_Y$ are now different from the one that applies to (\ref{defkd}). In (\ref{defkd-zperpx}), $k_D$ ($k_Y$) measures the relative explanatory power in explaining variation in treatment (outcome) of the residualized part of the unobserved confounder, $Z^{\perp X}$, in comparison to the covariate, $X_j$, chosen for benchmarking. Whether a comparison of the explanatory powers of $Z^{\perp X}$ and $X_j$ is feasible and meaningful will have to be decided by the researcher from the specific context of the study.
	
	To derive the exact expression for $R^2_{D \sim Z|X}$, we first note, using corollary~\ref{cor-partial-r2}, that $R^2_{D \sim Z|X} = R^2_{D \sim Z^{\perp X}|X}$. Hence, all we need to do is find an upper bound for $R^2_{D \sim Z^{\perp X}|X}$. But
	\[
	R^2_{D \sim Z^{\perp X}|X} = \frac{R^2_{D \sim X + Z^{\perp X}} - R^2_{D \sim X}}{1 - R^2_{D \sim X}} = \frac{R^2_{D \sim Z^{\perp X}}}{1 - R^2_{D \sim X}} = \frac{k_D R^2_{D \sim X_j}}{1-R^2_{D \sim X}} = k_D \left( \frac{R^2_{D \sim X_j}}{1-R^2_{D \sim X}}\right) 
	\]
	with $k_D$ defined in (\ref{defkd-zperpx}). Hence, we have an expression for $R^2_{D \sim Z^{\perp X}|X}$, and hence for $R^2_{D \sim Z|X}$. A similar argument gives us an expression for $R^2_{Y \sim Z|X}$,
	\[
	R^2_{Y \sim Z|X} = R^2_{Y \sim Z^{\perp X}|X} = k_Y \left( \frac{R^2_{Y \sim X_j}}{1-R^2_{Y \sim X}}\right),
	\]
	with $k_Y$ defined in (\ref{defkd-zperpx}).
	
	\subsubsection{Partial $R^2$ benchmarking using one covariate}\label{sec:partialr2-resid} 
	
	Note once again that the definition of $k_D$ and $k_Y$ has to use $Z^{\perp X}$, instead of $Z$. Thus, let
	\begin{equation}\label{kd-partial-single-perpx}
		k_D \coloneqq \frac{R^2_{D \sim Z^{\perp X}|X_{-j}}}{R^2_{D \sim X_j|X_{-j}}}, \quad k_Y \coloneqq \frac{R^2_{Y \sim Z^{\perp X}|X_{-j}}}{R^2_{D \sim X_j|X_{-j}}}.
	\end{equation}
	The interpretation of $k_D$ and $k_Y$ are, once again, very different from (\ref{kd-partial-single}) when $Z$ was assumed to be orthogonal to $X$. To interpret the definition of $k_D$ and $k_Y$ in (\ref{kd-partial-single-perpx}), let us use (\ref{def-partialr2}) to rewrite them:
	\[
	k_D \coloneqq \frac{R^2_{D \sim Z^{\perp X}+X_{-j}} - R^2_{D \sim X_{-j}}}{R^2_{D \sim X_j+X_{-j}} - R^2_{D \sim X_{-j}}}, \quad k_Y \coloneqq \frac{R^2_{Y \sim Z^{\perp X}+X_{-j}} - R^2_{Y \sim X_{-j}}}{R^2_{Y \sim X_j+X_{-j}} - R^2_{Y \sim X_{-j}}}.
	\]
	Now, we see that $k_D$ ($k_Y$) in (\ref{kd-partial-single-perpx}) measures the ratio of (a) the increment in total $R^2$ when $Z^{\perp X}$ (residualized unobserved confounder) is added to a regression of $D$ (or $Y$) on $X_{-j}$ (the set of covariates not used for benchmarking), and (b) the increment in total $R^2$ when $X_j$ (the covariate used for benchmarking) is added to a regression of $D$ ( or $Y$) on $X_{-j}$.\footnote{Since $Z^{\perp X} \perp X_{-j}$, the numerator becomes $R^2_{D \sim Z^{\perp X}}$ (or $R^2_{Y \sim Z^{\perp X}}$), the total $R^2$ in a regression of $D$ (or Y) on $Z^{\perp X}$.} This is the specific way in which $k_D$ ($k_Y$) now capture the relative explanatory power in explaining variation in treatment (outcome) of the residualized part of the unobserved confounder, $Z^{\perp X}$, in comparison to the covariate, $X_j$, chosen for benchmarking. Once again, whether a comparison of the explanatory powers of $Z^{\perp X}$ and $X_j$ in the particular way captured in (\ref{kd-partial-single-perpx}) is feasible and meaningful will have to be decided by the researcher from the specific context of the study.
	
	If we insert $Z^{\perp X}$ in place of $Z$ in the argument in \citet[appendix~B.2, page 35]{cinelli-hazlett-2020}, and use corollary~\ref{cor-partial-r2}, we get
	\[
	R^2_{D \sim Z|X} = R^2_{D \sim Z^{\perp X}|X} = k_D \left( \frac{R^2_{D \sim X_j|X_{-j}}}{1-R^2_{D \sim X_j|X_{-j}}}\right), 
	\]
	where $k_D$ is defined in (\ref{kd-partial-single-perpx}). A similar argument gives us 
	\[
	R^2_{Y \sim Z|X} = R^2_{Y \sim Z^{\perp X}|X} = k_Y \left( \frac{R^2_{Y \sim X_j|X_{-j}}}{1-R^2_{Y \sim X_j|X_{-j}}}\right), 
	\]
	where $k_Y$ is defined in (\ref{kd-partial-single-perpx}).
	
	\subsubsection{Partial $R^2$ benchmarking using multiple covariates}
	We now define
	\begin{equation}\label{kd-partial-multiple-perpx}
		k_D \coloneqq \frac{R^2_{D \sim Z^{\perp X}|X_{-(1,2, \ldots, j)}}}{R^2_{D \sim X_{(1,2 \ldots, j)}|X_{-(1,2, \ldots, j)}}}; \quad k_Y \coloneqq \frac{R^2_{Y \sim Z^{\perp X}|X_{-(1,2, \ldots, j)}}}{R^2_{Y \sim X_{(1,2 \ldots, j)}|X_{-(1,2, \ldots, j)}}}
	\end{equation}
	and note the change in interpretation. Writing $k_D$ and $k_Y$ using (\ref{def-partialr2}), we see that in (\ref{kd-partial-multiple-perpx}), $k_D$ (or $k_Y$) measures the ratio of (a) the increment in total $R^2$ when $Z^{\perp X}$ (residualized unobserved confounder) is added to a regression of $D$ (or $Y$) on $X_{-(1,2, \ldots, j)}$ (the set of covariates not used for benchmarking), and (b) the increment in total $R^2$ when $X_{(1,2, \ldots, j)}$ (the set of covariates used for benchmarking) is added to a regression of $D$ ( or$Y$) on $X_{-(1,2, \ldots, j)}$.\footnote{Since $Z^{\perp X} \perp X_{-(1,2, \ldots, j)}$, the numerator becomes $R^2_{D \sim Z^{\perp X}}$ (or $R^2_{Y \sim Z^{\perp X}}$), the total $R^2$ in a regression of $D$ (or Y) on $Z^{\perp X}$.} This is the specific way in which $k_D$ ($k_Y$) now capture the relative explanatory power in explaining variation in treatment (outcome) of the residualized part of the unobserved confounder, $Z^{\perp X}$, in comparison to the covariate, $X_j$, chosen for benchmarking. Once again, whether a comparison of the explanatory powers of $Z^{\perp X}$ and $X_j$ in the particular way captured in (\ref{kd-partial-single-perpx}) is feasible and meaningful will have to be decided by the researcher from the specific context of the study.
	
	If we insert $Z^{\perp X}$ in place of $Z$ in the argument in \citet[appendix~B.2, page 36]{cinelli-hazlett-2020}, and use corollary~\ref{cor-partial-r2}, we get  
	\[
	R^2_{D \sim Z|X} = R^2_{D \sim Z^{\perp X}|X} = k_D \left( \frac{R^2_{D \sim X_{(1,2, \ldots j)}|X_{-(1,2, \ldots, j)}}}{1-R^2_{D \sim X_{(1,2, \ldots j)}|X_{-(1,2, \ldots, j)}}}\right),
	\]
	with $k_D$ defined in (\ref{kd-partial-multiple-perpx}). A similar argument gives us
	\[
	R^2_{Y \sim Z|X} = R^2_{Y \sim Z^{\perp X}|X} = k_Y \left( \frac{R^2_{Y \sim X_{(1,2, \ldots j)}|X_{-(1,2, \ldots, j)}}}{1-R^2_{Y \sim X_{(1,2, \ldots j)}|X_{-(1,2, \ldots, j)}}}\right),
	\]
	with $k_Y$ defined in (\ref{kd-partial-multiple-perpx}).
	
	\section{Covariate benchmarking with orthogonality: Second step}\label{sec:cov-bench-ortho-2}
	The first step of covariate benchmarking gives us expressions for $R^2_{D \sim Z|X}$ and $R^2_{Y \sim Z|X}$. In the second step, we have to find an expression for $R^2_{Y \sim Z|D,X}$. As we have seen above in section~\ref{sec:cov-bench-general}, this will involve finding the exact magnitude of $\left| R_{Y \sim Z|X} - R_{Y \sim D|X} R_{D \sim Z|X} \right|$, and for this purpose I need the next result.

	\begin{proposition}\label{lem2}
		For any three non-zero real numbers, $a, b, c$, the following two statements are equivalent. 
		\begin{enumerate}
			\item $\sign\left( a-bc\right) = \sign\left( bc\right) $; 
			\item $|a-bc|=|a| - |bc|$.
		\end{enumerate}
	\end{proposition}

	We know that for any three real numbers, $a,b$, and $c$, we have $|a-bc| \geq |a| - |bc|$. Using proposition~\ref{lem2}, now we get another result that I will use below:
	\begin{equation}\label{sign-diff}
		\sign\left( a-bc\right) \neq \sign\left( bc\right)  \iff |a-bc| > |a| - |bc|.
	\end{equation}
	 
	\subsection{Confounder reduces magnitude of treatment effect}
	We first consider the case where the unobserved confounder reduces the magnitude of the true treatment effect towards zero. We can capture this case with the condition that the absolute value of the treatment effect from the restricted model is strictly less than the treatment effect from the full model: $|\hat{\tau}_{\text{res}}|<|\hat{\tau}|$. We need to consider two sub-cases: (a) signs of the estimated and true treatment effect are same; (b) signs of the estimated and true treatment effect are opposite. 
	
	\subsubsection{Signs of the estimated and true treatment effect are same}
	In this case, we have
	\begin{equation}
		\frac{\widehat{\text{bias}}}{\hat{\tau}_{\text{res}}} <0.
	\end{equation} 
	To see this consider two cases: $\hat{\tau}>0$ or $\hat{\tau} < 0$. If $\hat{\tau}>0$, then $\hat{\tau}_{\text{res}} \in \left( -\hat{\tau}, \hat{\tau}\right) $. Hence, using ($\ref{def-bias}$), we see that $\widehat{\text{bias}}<0$. Since the signs of the estimated and true treatment effect are same, $\hat{\tau}_{\text{res}}>0$. Hence, $\frac{\widehat{\text{bias}}}{\hat{\tau}_{\text{res}}} <0$. On the other hand, if $\hat{\tau}\leq 0$, then $\hat{\tau}_{\text{res}} \in \left( \hat{\tau}, -\hat{\tau}\right) $, so that $\widehat{\text{bias}}>0$. Since the signs of the estimated and true treatment effect are same, $\hat{\tau}_{\text{res}}<0$. Hence, $\frac{\widehat{\text{bias}}}{\hat{\tau}_{\text{res}}} <0$. Now, I will use (\ref{bias-rel}) and proposition~\ref{lem2} to investigate the magnitude of $\left| R_{Y \sim Z|X} - R_{Y \sim D|X} R_{D \sim Z|X} \right|$. We need to consider two cases.
	
	\textit{Case~1:} $R_{Y \sim D|X}>0$. Since $\frac{\widehat{\text{bias}}}{\hat{\tau}_{\text{res}}} <0$, using (\ref{bias-rel}), we see that the signs of $R_{Y \sim Z|D,X}$ and $ R_{D \sim Z|X}$ are opposite. Suppose $R_{Y \sim Z|D,X}>0$ and $ R_{D \sim Z|X}<0$; then $R_{Y \sim D|X}R_{D \sim Z|X}<0$. Thus, we see that the signs of $R_{Y \sim Z|D,X}$ and $R_{Y \sim D|X}R_{D \sim Z|X}$ are opposite. Letting $a=R_{Y \sim Z|D,X}, b=R_{Y \sim D|X}$ and $c=R_{D \sim Z|X}$ in (\ref{sign-diff}) shows that 
	\[
	\left| R_{Y \sim Z|X} - R_{Y \sim D|X} R_{D \sim Z|X} \right|>\left| R_{Y \sim Z|X} \right| - \left| R_{Y \sim D|X} R_{D \sim Z|X} \right|.
	\] Thus we do not have an exact expression or an upper bound for $\left| R_{Y \sim Z|X} - R_{Y \sim D|X} R_{D \sim Z|X} \right|$. A similar argument shows that if $R_{Y \sim Z|D,X}<0$ and $ R_{D \sim Z|X}>0$, then also we do not have an upper bound or an exact expression for $\left| R_{Y \sim Z|X} - R_{Y \sim D|X} R_{D \sim Z|X} \right|$. 

	\textit{Case~2:} $R_{Y \sim D|X}<0$. Since $\frac{\widehat{\text{bias}}}{\hat{\tau}_{\text{res}}} <0$, using (\ref{bias-rel}), we see that the signs of $R_{Y \sim Z|D,X}$ and $ R_{D \sim Z|X}$ are the same. Suppose $R_{Y \sim Z|D,X}>0$ and $ R_{D \sim Z|X}>0$; then $R_{Y \sim D|X}R_{D \sim Z|X}<0$. Thus, we see that the signs of $R_{Y \sim Z|D,X}$ and $R_{Y \sim D|X}R_{D \sim Z|X}$ are opposite. Hence, we do not have an exact expression or an upper bound for $\left| R_{Y \sim Z|X} - R_{Y \sim D|X} R_{D \sim Z|X} \right|$. A similar argument gives the same result when $R_{Y \sim Z|D,X}<0$ and $ R_{D \sim Z|X}<0$.

	\subsubsection{Signs of the estimated and true treatment effect are different}\label{sec:confd-reduce-same-sign}
	In this case, we have
	\begin{equation}
		\frac{\widehat{\text{bias}}}{\hat{\tau}_{\text{res}}} >0.
	\end{equation} 
	To see this consider, once again, two cases: $\hat{\tau}>0$ or $\hat{\tau} < 0$. If $\hat{\tau}>0$, then $\hat{\tau}_{\text{res}} \in \left( -\hat{\tau}, \hat{\tau}\right) $. Hence, using ($\ref{def-bias}$), we see that $\widehat{\text{bias}}<0$. Since the signs of the estimated and true treatment effect are opposite, $\hat{\tau}_{\text{res}}<0$. Hence, $\frac{\widehat{\text{bias}}}{\hat{\tau}_{\text{res}}} >0$. On the other hand, if $\hat{\tau}< 0$, then $\hat{\tau}_{\text{res}} \in \left( \hat{\tau}, -\hat{\tau}\right) $, so that $\widehat{\text{bias}}>0$. Since the signs of the estimated and true treatment effect are opposite, $\hat{\tau}_{\text{res}}>0$. Hence, $\frac{\widehat{\text{bias}}}{\hat{\tau}_{\text{res}}} >0$. We need to, once again, consider two cases.
	
	\textit{Case~1:} $R_{Y \sim D|X}>0$. Since $\frac{\widehat{\text{bias}}}{\hat{\tau}_{\text{res}}} >0$, using (\ref{bias-rel}), we see that the signs of $R_{Y \sim Z|D,X}$ and $ R_{D \sim Z|X}$ are the same. Suppose $R_{Y \sim Z|D,X}>0$ and $ R_{D \sim Z|X}>0$; then $R_{Y \sim D|X}R_{D \sim Z|X}>0$. Thus, we see that the signs of $R_{Y \sim Z|D,X}$ and $R_{Y \sim D|X}R_{D \sim Z|X}$ are the same. Letting $a=R_{Y \sim Z|D,X}, b=R_{Y \sim D|X}$ and $c=R_{D \sim Z|X}$ in proposition~\ref{lem2} gives us
	\[
	\left| R_{Y \sim Z|X} - R_{Y \sim D|X} R_{D \sim Z|X} \right|=\left| R_{Y \sim Z|X} \right| - \left| R_{Y \sim D|X} R_{D \sim Z|X} \right|.
	\] 
	Thus we have an exact expression for $\left| R_{Y \sim Z|X} - R_{Y \sim D|X} R_{D \sim Z|X} \right|$. A similar argument shows that if $R_{Y \sim Z|D,X}<0$ and $ R_{D \sim Z|X}<0$, then also we have the same expression for $\left| R_{Y \sim Z|X} - R_{Y \sim D|X} R_{D \sim Z|X} \right|$. 
	
	\textit{Case~2:} $R_{Y \sim D|X}<0$. Since $\frac{\widehat{\text{bias}}}{\hat{\tau}_{\text{res}}} >0$, using (\ref{bias-rel}), we see that the signs of $R_{Y \sim Z|D,X}$ and $ R_{D \sim Z|X}$ are opposite. Suppose $R_{Y \sim Z|D,X}>0$ and $ R_{D \sim Z|X}<0$; then $R_{Y \sim D|X}R_{D \sim Z|X}>0$. Thus, we see that the signs of $R_{Y \sim Z|D,X}$ and $R_{Y \sim D|X}R_{D \sim Z|X}$ are the same. Letting $a=R_{Y \sim Z|D,X}, b=R_{Y \sim D|X}$ and $c=R_{D \sim Z|X}$ in proposition~\ref{lem2} gives us
	\[
	\left| R_{Y \sim Z|X} - R_{Y \sim D|X} R_{D \sim Z|X} \right|=\left| R_{Y \sim Z|X} \right| - \left| R_{Y \sim D|X} R_{D \sim Z|X} \right|.
	\] 
	Thus we have an exact expression for $\left| R_{Y \sim Z|X} - R_{Y \sim D|X} R_{D \sim Z|X} \right|$. A similar argument gives the same result when $R_{Y \sim Z|D,X}<0$ and $ R_{D \sim Z|X}>0$.
	
	\subsection{Confounder increases magnitude of estimate}\label{sec:confd-increase}
	Now we consider the case where the unobserved confounder increases the magnitude of the true treatment effect away from zero. We can capture this case with the condition that the absolute value of the treatment effect from the restricted model is strictly greater than the treatment effect from the full model: $|\hat{\tau}_{\text{res}}|>|\hat{\tau}|$. Just as in the previous case, we need to consider two sub-cases: (a) signs of the estimated and true treatment effect are same; (b) signs of the estimated and true treatment effect are opposite.\footnote{For details see appendix~\ref{sec:app-confd-increase}.} Using similar arguments as in the above case, we can see that, in both the sub-cases,
	\begin{equation}
		\frac{\widehat{\text{bias}}}{\hat{\tau}_{\text{res}}} >0.
	\end{equation}
	Once again, using similar arguments as in section~\ref{sec:confd-reduce-same-sign}, we have
	\[
	\left| R_{Y \sim Z|X} - R_{Y \sim D|X} R_{D \sim Z|X} \right|=\left| R_{Y \sim Z|X} \right| - \left| R_{Y \sim D|X} R_{D \sim Z|X} \right|.
	\]
	
	\subsection{Summary}
	In this section, I have investigated scenarios when we have 
	\[
	\left| R_{Y \sim Z|X} - R_{Y \sim D|X} R_{D \sim Z|X}\right|  = \left| R_{Y \sim Z|X}\right|  - \left| R_{Y \sim D|X} R_{D \sim Z|X}\right|. 
	\]
	This is important because, in these cases, we can use (\ref{partialr2-rec}) to determine $|R_{Y \sim Z|D,X}|$ as 
	\[
	\left| R_{Y \sim Z|D,X} \right| = \frac{\left| R_{Y \sim Z|X}\right|  - \left| R_{Y \sim D|X}\right| \left|  R_{D \sim Z|X}\right|  }{\sqrt{1-R^2_{Y \sim D|X}} \sqrt{1-R^2_{D \sim Z|X}}},
	\]
	with information generated in the first step about $|R_{Y \sim Z|X}|$ and $|R_{D \sim Z|X}|$. This will then allow us to complete the second step of the sensitivity analysis. Let me summarize what I found.
	
	If the confounder increases the magnitude of the treatment effect away from zero, i.e. the estimated treatment effect is larger in absolute value than the true treatment effect, then we have an exact expression for $|R_{Y \sim Z|X} - R_{Y \sim D|X} R_{D \sim Z|X}|$ irrespective of the signs of the estimated and true treatment effect. Hence, we can use (\ref{partialr2-rec}) to determine $|R_{Y \sim Z|D,X}|$ unambiguously.
	
	If the confounder reduces the magnitude of the treatment effect towards zero, i.e. the estimated treatment effect is smaller in absolute value than the true treatment effect, then we need to consider the signs of the true and the estimated treatment effects. If the true and estimated treatment effects have opposite signs, then, once again, we have an exact expression for $|R_{Y \sim Z|X} - R_{Y \sim D|X} R_{D \sim Z|X}|$. If, on the other hand, the true and estimated treatment effects have same signs, then we do not have an exact expression for $|R_{Y \sim Z|X} - R_{Y \sim D|X} R_{D \sim Z|X}|$; instead we have a lower bound. Hence, we cannot find bounds for the omitted variable bias using the methodology in \citet{cinelli-hazlett-2020}.
	
	This last scenario is of course a special case where we can avoid an elaborate sensitivity analysis altogether. Since the confounder reduces the magnitude of the treatment effect towards zero without changing sign, we either have $0 < \hat{\tau}_{\text{res}} < \hat{\tau}$ or we have $ \hat{\tau} < \hat{\tau}_{\text{res}} < 0$. If the researcher can argue plausibly, e.g. on the basis of domain knowledge, that the true treatment effect is positive, then she can use the former case, i.e. $0 < \hat{\tau}_{\text{res}} < \hat{\tau}$, to assert that the estimated treatment effect, $ \hat{\tau}_{\text{res}} $, is a lower bound for the true treatment effect. If, on the other hand, the researcher can make a plausible case that the true treatment effect is negative, then she can use the latter case, i.e. $ \hat{\tau} < \hat{\tau}_{\text{res}} < 0$, to argue that the estimated treatment effect, $ \hat{\tau}_{\text{res}} $, is an upper bound of the true treatment effect.
	
	The upshot is that researchers can rest assured that the method for determining $|R_{Y \sim Z|D,X}|$ using (\ref{partialr2-rec}) covers all cases that matter. In the only case where (\ref{partialr2-rec}) cannot be used to determine $|R_{Y \sim Z|D,X}|$, i.e. the case where the confounder reduces the magnitude of the treatment effect towards zero without changing sign, the researcher can avoid conducting sensitivity analysis. Of course, in practice, it is never easy to know, for a specific study, which case a researcher faces. Hence, having all relevant cases covered is reassuring.

	\section{Some examples}\label{sec:examples}
		
	In Table~\ref{table:examples}, I report results of sensitivity analyses for some existing observational studies to highlight some of the interpretational issues that need to be kept in mind. The first row of table~\ref{table:examples} discusses the running example used in \citet[section~2]{cinelli-hazlett-2020}. The substantive issue under investigation is whether, in the context of the civil war in Darfur, exposure to violence (treatment) has a causal effect on the attitudes towards peace (outcome). In rows 2 to 6, I discuss the observational studies reported in \citet[Section~4.2]{oster-2019}. The substantive issue under investigation in these studies is the impact of maternal behavior on child outcomes, with a child's standardized IQ score as the outcome variable in rows 2 through 4, and a child's birth weight as the outcome variable in rows 5 and 6. 
	
	\subsection{Impact of exposure to violence on attitude towards peace}
	
	Political scientists are often interested in understanding the effect of exposure to violence on attitudes towards peace.\footnote{For further details, see \citet[section~2]{cinelli-hazlett-2020}. I accessed the data set used for the analysis in the R package \texttt{sensemakr}.} In the context of the civil war in Darfur, this question attained especial importance. To investigate this question, a researcher estimates the following regression model with individual-level data: $\text{PeaceIndex}$ is regressed on $\text{DirectHarm}$, along with control variables, where $\text{PeaceIndex}$ is an index to measure attitudes towards peace efforts, $\text{DirectHarm}$ measures the exposure to violence. The control variables are: $\text{Female}$ (gender of the individual), age, whether the individual was a farmer, herder, merchant or trader, household size, whether or not the individual voted in the past, and village-level fixed effects. Columns~1 and 2 of row~1 in table~\ref{table:examples} show that the treatment effect is $0.097$ and highly statistically significant.
	
	It is suspected that there is at least one important unobserved confounder, $Z$, (e.g. wealth of the individual) that is correlated both with exposure to violence (treatment) and attitude towards peace (outcome). The researcher does not have data on wealth of individuals. Hence, the researcher wishes to conduct sensitivity analysis regarding the possible effect of this omitted variable. Suppose, finally, the researcher knows, on the bases of domain knowledge, that gender of the individual is one of the, if not the, most important variables determining treatment (exposure to violence measured by $\text{DirectHarm}$). Hence, she chooses to use this variable for formal covariate benchmarking. 
	
	The first question of interest in the sensitivity analysis reported in row~1, table~\ref{table:examples}, is whether taking account of omitted variable bias can change the point estimate of the treatment effect to zero. Using a total $R^2$-based comparison, we see that $R^2_{D \sim Z|X}=0.268$ (column~6), $R^2_{Y \sim Z|D,X}=25.907$ (column~5), and $RV=13.878$ (column~3). Hence,  $\max \left\lbrace R^2_{D \sim Z|X}, R^2_{Y \sim Z|D,X}\%\right\rbrace > RV\%$. Thus, it is not possible to rule out that taking account of omitted variable bias can change the treatment effect to zero. If, on the other hand, the researcher uses a partial $R^2$-based comparison, we see that $R^2_{D \sim Z|X}=0.916$ (column~6), and $R^2_{Y \sim Z|D,X}=12.464$ (column~5). Hence,  $\max \left\lbrace R^2_{D \sim Z|X}, R^2_{Y \sim Z|D,X}\%\right\rbrace < RV\%$, and it is possible to conclude that omitted variable bias cannot reduce the treatment effect to zero.
	
	The second, related, question is whether we can conclude that the treatment effect is different from zero even after we take account of the uncertainty of estimation, i.e. see whether zero is contained in the $95\%$ confidence interval around the estimated treatment effect. To answer this question we need to compare $\max \left\lbrace R^2_{D \sim Z|X}, R^2_{Y \sim Z|D,X}\%\right\rbrace$ and $RV_{\alpha}$ (column~4). Both the total $R^2$-based comparison and the partial $R^2$-based comparison lead to the conclusion that $\max \left\lbrace R^2_{D \sim Z|X}, R^2_{Y \sim Z|D,X}\%\right\rbrace >RV_{\alpha}$. Hence, it is not possible to rule out a zero treatment effect once we take account of the uncertainty of estimation. 
	
	Finally, we can consider the extreme scenario where the unobserved confounder would explain all the residuals variation of the outcome. Would such a confounder manage to eliminate the treatment effect? To answer this question, we need to compare $R^2_{D \sim Z|X}$ (6 or 8) and $R^2_{Y \sim D|X}$ (column~9). For both the total $R^2$-based comparison and the partial $R^2$-based comparison, $R^2_{D \sim Z|X}<R^2_{Y \sim D|X}$. Hence, we can conclude that such a confounder will not be able to completely wipe out the treatment effect.
	
	\subsection{Impact of maternal behavior on child outcomes}
	
	In the studies of the effect of maternal behavior on child outcome for the U.S. using NLSY data (rows~2 through 6 in table~\ref{table:examples}), two child outcomes are studied: a child's standardized IQ score (rows~2 through 4 in table~\ref{table:examples}) and a child's birth weight (rows~5 and 6 in table~\ref{table:examples}). In the study of child IQ, three treatment variables are used in turn: months of breastfeeding (row~2), any drinking of alcohol in pregnancy (row~3), and an indicator for being low birth-weight and preterm (row~4). In studying child birth-weight, two treatment variables are used, one by one: maternal smoking during pregnancy (row~5), and maternal drinking during pregnancy (row~6). The following control variables are used for both studies: child race, maternal age, maternal education, maternal income, maternal marital status. The question of interest is whether the treatment variables, each on their own, have any causal impact on the outcome variables.\footnote{For further details, see \citet[Section~4.2]{oster-2019}. I accessed the data sets used for the analysis from Professor Emily Oster's website: \url{https://drive.google.com/file/d/0B1U4uS7GkkxbV0VkZmd0ZVlDVDA/view?usp=sharing}} 
	
	In answering each of the questions about the impact of maternal behavior on child outcomes using observational studies, researchers face the problem of omitted variables ``largely associated with omitted socioeconomic status and family background.'' \citep[p.~198]{oster-2019}. Since mother's years of schooling is one of the most important determinants of treatment, and can also be a good proxy for socioeconomic status, we use this variable for formal covariate benchmarking in these studies. Sensitivity analysis presented in rows~2 through 6 in table~\ref{table:examples} show the following: 
	\begin{itemize}
		\item Row~2, table~\ref{table:examples}: Maternal breastfeeding has a positive impact ($0.017$) on a child's IQ score, and this positive impact will not be reduced to zero even if we take account of omitted variable bias because $R^2_{D \sim Z|X}$ and $ R^2_{Y \sim Z|D,X}$ are both less than $RV$, irrespective of whether we use total or partial $R^2$-based comparison for covariate benchmarking. Once we take account of estimation uncertainty, the conclusion changes. The adjusted confidence interval does not contain zero if we use a partial $R^2$-based comparison, but it does contain zero if we use a partial $R^2$-based comparison: $R^2_{Y \sim Z|D,X}$ (column~5) is greater than $RV_{\alpha}$ (column~4). Turning to the extreme or worst-case confounder scenario, we see that $R^2_{D \sim Z|X}$ is smaller than $R^2_{Y \sim D|X}$ for both total and partial $R^2$-based comparisons. This shows that a worst-case confounder that explains all residual variance of the outcome would still not be able to explain away the result.
		
		\item Row~3, table~\ref{table:examples}: Drinking by the mother during pregnancy has a positive impact ($ 0.050 $) on a child's IQ score. But this counterintuitive positive estimate is completely wiped out once we take account of omitted variable bias: $R^2_{Y \sim Z|D,X}$ for total $R^2$-based comparison (column~5) and for partial $R^2$-based comparison (column~7) are both larger than $RV$ (column~3). Since this result holds for the point estimate, we do not even need to look at the result about the confidence interval or conduct extreme confounder analysis.
		
		\item Row~4, table~\ref{table:examples}: Being low birth-weight and preterm has a negative impact (-$ 0.125 $) on a child's IQ score. As in the previous row's result, the negative estimate of the treatment effect is completely wiped out once we take account of omitted variable bias: $R^2_{Y \sim Z|D,X}$ for total $R^2$-based comparison (column~5) and for partial $R^2$-based comparison (column~7) are both larger than $RV$ (column~3). Once again, since this result holds for the point estimate, we do not need to look at the result about the confidence interval  or conduct extreme confounder analysis.
		
		\item Row~5, table~\ref{table:examples}: Smoking by the mother during pregnancy has a negative impact (-$ 172.51 $) on a child's birth-weight. The negative estimate remains intact even after we take account of omitted variable bias, irrespective of whether we use total or partial $R^2$-based covariate benchmarking. To see this, note that $RV_{\alpha}$ ($12.107$, column~4) is greater than $R^2_{D \sim Z|X}$ and $ R^2_{Y \sim Z|D,X}$ for both total and partial $R^2$-based comparisons. Hence, not only is the point estimate different from zero, the $95\%$ confidence interval does not contain zero. Thus, even after taking account of estimation uncertainty, we can assert that omitted variable bias cannot overturn the results. What does the worst-case confounder analysis show? Here we see mixed results. If we use a partial $R^2$-based comparison, then $R^2_{D \sim Z|X}$ (column~8) is less than $R^2_{Y \sim D|X}$ (column~9); if, on the other hand, we use a total $R^2$-based comparison, then $R^2_{D \sim Z|X}$ (column~6) is larger than $R^2_{Y \sim D|X}$ (column~9).
			
		\item Row~6, table~\ref{table:examples}: Drinking by the mother during pregnancy has a negative impact (-$ 14.14 $) on a child's birth-weight. The negative estimate of the treatment effect remains intact even after we take account of omitted variable bias, irrespective of whether we use total or partial $R^2$-based covariate benchmarking. To see this, note that $RV_{\alpha}$ ($0.972$, column~4) is greater than $R^2_{D \sim Z|X}$ and $ R^2_{Y \sim Z|D,X}$ for both total and partial $R^2$-based comparisons. Thus, not only is the point estimate different from zero, even the $95\%$ confidence interval does not contain zero. We can conclude that even after taking account of estimation uncertainty, omitted variable bias cannot overturn the results. What does the worst-case confounder analysis show? We see that $R^2_{D \sim Z|X}$ is smaller than $R^2_{Y \sim D|X}$ for both total and partial $R^2$-based comparisons. Thus, a worst-case confounder that explains all residual variance of the outcome would still not be able to explain away the result.
		
	\end{itemize} 
		
	\begin{table}[hbt!]
		\begin{center}
			\begin{threeparttable}
				\caption{Sensitivity Analysis of Several Observational Studies}
				\label{table:examples}
				\begin{tabular}{lccccccccc}
					\toprule
					&  &  &  &  & \multicolumn{2}{c}{Total $R^2$ comparison} & \multicolumn{2}{c}{Partial $R^2$ comparison} &  \\
					\cline{6-7} \cline{8-9}
					& Est. & SE & $ RV_q $ & $ RV_{q,\alpha} $ & $R^2_{Y \sim Z|D,X}$ & $R^2_{D \sim Z|X}$ & $R^2_{Y \sim Z|D,X}$ & $R^2_{D \sim Z|X}$ & $R^2_{Y \sim D|X}$ \\
					&  &  & (\%) & (\%) & (\%) & (\%) & (\%) & (\%) & (\%) \\
					\cline{2-10} 
					& (1) & (2) & (3) & (4) & (5) & (6) & (7) & (8) & (9) \\ 
					\hline
					PC:VL & 0.097 & 0.023 & 13.878 & 7.626 & 25.907 & 0.268 & 12.464 & 0.916 & 2.187 \\ 
					IQ:BF & 0.017 & 0.002 & 8.497 & 6.247 & 7.733 & 3.134 & 3.513 & 0.825 & 0.783 \\ 
					IQ:DR & 0.050 & 0.023 & 2.676 & 0.284 & 7.085 & 0.243 & 3.341 & 0.195 & 0.074 \\ 
					IQ:LB & -0.125 & 0.050 & 3.148 & 0.699 & 7.053 & 0.007 & 3.501 & 0.021 & 0.102 \\ 
					BW:SM & -172.510 & 13.285 & 14.098 & 12.107 & 0.156 & 2.929 & 0.003 & 1.864 & 2.261 \\ 
					BW:DR & -14.149 & 5.065 & 3.221 & 0.972 & 0.411 & 0.003 & 0.067 & 0.033 & 0.035 \\ 
					
					\bottomrule
				\end{tabular}
				\begin{tablenotes}
					\small
					\item \textit{Notes:} The first row of this table reports sensitivity analysis for the running example used in \citet{cinelli-hazlett-2020}; the last five rows report sensitivity analyses of the five models reported in \citet[Table 3]{oster-2019}. Row~1: effect of exposure to violence (VL) on attitudes towards peace (PC); Row~1: effect of exposure to violence (VL) on attitudes towards peace (PC); Row~2: effect of maternal breastfeeding (BF) on child's IQ score (IQ); Row~3: effect of maternal drinking of alcohol during pregnancy (DR) on child's IQ score (IQ); Row~4: effect of low birth weight and preterm (LB) on child's IQ score (IQ); Row~5: effect of maternal smoking during pregnancy (SM) on child's birth weight (BW); Row~6: effect of maternal drinking of alcohol during pregnancy (DR) on child's birth weight (BW). Est. = estimate; SE = standard error; $ RV $ and $ RV_{\alpha} $ are the robustness values discussed in section~\ref{sec:step1}. For a discussion of total $R^2$-based comparison, see section~\ref{sec:totalr2-resid}; for a discussion of partial $R^2$-based comparisons, see section~\ref{sec:partialr2-resid}; for discussion of $R^2_{Y \sim Z|D,X}$, see section~\ref{sec:cov-bench-ortho-2}. For row~1, the benchmark covariate is female (gender of the individual); for rows~2--6, the benchmark covariate is years of schooling of mother. In all studies, $k_D=1$, $k_Y=1$, $q=1$ (taking account of omitted variable bias can reduce treatment effect to zero) and $\alpha=0.05$ (where relevant, $95\%$ confidence intervals are considered). 
				\end{tablenotes}
			\end{threeparttable}
		\end{center}
		
	\end{table}
	
	\subsection{Two issues to keep in mind}
	The examples discussed above raise two issues that need to be noted. First, the overall conclusion of sensitivity analysis about omitted variable bias depends on whether the researcher chooses to use a total $R^2$-based or a partial $R^2$-based formal covariate benchmarking. We saw this issue come up in many cases, e.g. row~1 (the impact of exposure to violence on the attitudes towards peace), row~2 (the impact of maternal breastfeeding on child IQ score) and the extreme confounder analysis in row~5 (effect of smoking by mother during pregnancy on child's birth-weight) in table~\ref{table:examples}. Of course, this is not surprising. After all total $R^2$-based or a partial $R^2$-based formal covariate benchmarking rely on different concepts of `relative importance' of the unobserved confounder and the observed covariates. In the former a ratio of total $R^2$ is used; in the latter a ratio of partial $R^2$ is used. \citet[p.~2]{kruskal-majors-1989} has pointed out, on the basis of a survey of many studies, that different concepts of relative importance often affect conclusions of research, and that is what we have found. The happy outcome is, of course, the one where both total $R^2$-based or a partial $R^2$-based formal covariate benchmarking lead to the same conclusion. If the conclusions differ, then researchers will need to justify using one or the other.
	
	Second, the researcher needs to be aware of the different interpretations or comparisons that are entailed by the total $R^2$-based versus the partial $R^2$-based formal covariate benchmarking. As an example, let us look at this issue of interpretation in the case of the study of the impact of exposure to violence on the attitudes towards peace. If the researcher decided to use total $R^2$-based covariate benchmarking, she will interpret the parameters, $k_D$ and $k_Y$, as follows: $k_D$ ($k_Y$) measures the relative explanatory power in explaining variation in exposure to violence (or, attitude towards peace) of the residualized part of the unobserved confounder, $Z^{\perp X}$, in comparison to the covariate, `Female'. In this case, the residualized part of the unobserved confounder is the part of $\text{Wealth}$ that is not explained linearly by gender, age, whether the individual was a farmer, herder, merchant or trader, the household size of the individual and whether or not the individual voted in the past, and village fixed effects. Since the researcher has to reason about the explanatory power of the residual, and not the unobserved covariate itself, she faces some difficulty. How can we reason about the part of `Wealth' that is not explained linearly by the observed covariates? 
	
	If, on the other hand, the researcher decides to use partial $R^2$-based covariate benchmarking, she will have to interpret $k_D$ and $k_Y$ differently. In this case, $k_D$ ($k_Y$) measures the ratio of two things: (a) the increment in total $R^2$ when the residualized unobserved confounder is added to a regression of exposure to violence (or attitude towards peace) on all the observed covariates other than `Female,' and (b) the increment in total $R^2$ when Female is added to a regression of exposure to violence (or attitude towards peace) on all the observed covariates other than Female. The researcher faces the same difficulty as in the total $R^2$-based comparisons. She has to, once again, reason about the part of `Wealth' that is not explained linearly by the observed covariates. Is there a reliable and intuitive way to do this?

	\section{Conclusion and limitations}\label{sec:conclusion}

	In an innovative and important contribution to the literature on omitted variable bias, \citet{cinelli-hazlett-2020} have proposed a methodology for conducting sensitivity analysis using partial $R^2$ measures. In their proposed methodology, the key step of formal covariate benchmarking---to generate upper bounds for measures of association between the unobserved confounder and the treatment (and outcome)---relies on the assumption that either the unobserved confounder is orthogonal to the set of observed covariates or that the analysis is conducted with the residualized part of the unobserved confounder. In this paper I have explained the need for these orthogonality assumptions and clarified a critical step in the computation of the sensitivity parameter $R^2_{Y \sim Z|D,X}$ (partial $R^2$ of the outcome with the unobserved confounder, conditioning on the treatment and the set of observed covariates). 
	
	I would like to conclude with some observations about the limitations of sensitivity analysis and the inherent difficulty of dealing with omitted variable bias in the context of observational studies. 
	
	Recall that operationalizing the standard expression of omitted variable bias involves knowing the signs of partial correlations of the unobserved confounder with the outcome and with the treatment. In many situations, e.g. when the unobserved confounder is not a single but an index of multiple variables, it is not possible to know these signs with any degree of certainty. In other cases, only knowing the signs are not enough, and researchers need to reason about their magnitudes too. This provides a justification for sensitivity analysis, i.e. knowing how strong the magnitudes of the partial correlations (or other similar measures of association) need to be to overturn the results of observational studies.
	
	Sensitivity analysis quantifies measures of association between the unobserved confounder with the outcome, and with the treatment, that can be deemed problematic (in the sense that they can overturn the results of observational studies); let us call these measures of association as robustness values (the parameters $RV_q$ and $RV_{q,\alpha}$ used in \citet{cinelli-hazlett-2020} are examples). Notice that the difficulty has not been completely addressed. This is because, in specific studies, researchers will then need to form judgments about whether the robustness values are too large or too small. How will they do so? After all the confounder is unobserved. Formal and informal covariate benchmarking steps in to help. 
	
	Using information about measures of association between an observed covariate (or a set of observed covariates) with the treatment and with the outcome, researchers try to find bounds for the measures of association between the unobserved confounder with the outcome and with the treatment. Notice that the difficulty has still not been fully resolved. This is because researchers now need to reason about the explanatory power of the unobserved confounder and compare that to the explanatory power of the benchmark covariate (the parameters $k_D$ and $k_Y$ used in \citet{cinelli-hazlett-2020} are examples of parameters whose construction requires researchers to make such comparisons). Difficult as this is, it turns out that even this is not sufficient. 
	
	In fact, we need to make these comparisons of explanatory powers not with respect to the unobserved confounder but rather with respect to the residualized part of the unobserved confounder, i.e. the part of the unobserved confounder that is not linearly explained by the set of observed covariates. It seems far from straightforward how one can reliably reason about the residualized part of the unobserved confounder and assess its explanatory power in comparison to some other covariate (or set of covariates) because, first, the confounder is unobserved, and second, it is not easy to pin down the part of the confounder that remains once we subtract a linear function of the set of observed covariates.

	\bibliographystyle{apalike}
	\bibliography{endocon_refs}
	
	\begin{appendices}
		
		\section{Proofs}
		\subsection{Proof of Lemma~\ref{lemma:decomp}}
		\begin{proof}
			Using results on the inverse of partitioned matrices, it can be shown \citep[page~323]{rao-etal-2008} that
			\begin{equation}
				P_W = P_X + \frac{ \left( I-P_X\right) Z Z'\left( I-P_X\right)  }{Z'\left( I-P_X\right)Z}. 
			\end{equation}
			Using the definition of $Z^{\perp X}$, we see that
			\begin{equation}
				\frac{ \left( I-P_X\right) Z Z'\left( I-P_X\right)  }{Z'\left( I-P_X\right)z}  = P_{Z^{\perp X}},
			\end{equation}
			where I use the fact that $\left( I-P_X\right)$ is also a projection matrix (onto the orthogonal complement of the column space of $X$) and hence symmetric and idempotent.
		\end{proof}
		
		\subsection{Proof of Theorem~\ref{thm:r2-decomp}}
		\begin{proof}
			Using lemma~\ref{lemma:decomp}, we have
			\[
			P_W M^0 P_W = P_W M^0 M^0 P_W = \left( M^0 P_X + M^0 P_{Z^{\perp X}}\right)'\left( M^0 P_X + M^0 P_{Z^{\perp X}} \right).  
			\]
			This becomes		
			\begin{equation}\label{eq:decomp1}
				P_W M^0 P_W = P_X M^0 P_X + P_{Z^{\perp X}} M^0 P_{Z^{\perp X}}
			\end{equation}
			because the cross product terms on the extreme right hand side is
			\[
			P_X M^0 M^0 P_{Z^{\perp X}} = P_X M^0 P_{Z^{\perp X}} = P_X P_{Z^{\perp X}} = 0.
			\]
			Note that the penultimate equality is true because
			\begin{align*}
				M^0 P_{Z^{\perp X}} & = M^0 Z^{\perp X} \left[  (Z^{\perp X})' Z^{\perp X} \right] ^{-1} (Z^{\perp X})' \\
				& = Z^{\perp X} \left[  (Z^{\perp X})' Z^{\perp X} \right] ^{-1} (Z^{\perp X})' \\
				& = P_{Z^{\perp X}}
			\end{align*}		
			where, because $ Z^{\perp X} $ is a regression residual vector, we have $M^0 Z^{\perp X} = Z^{\perp X}$ \citep[page~40]{greene}. The final equality is true because 
			\[
			P_X P_{Z^{\perp X}}  = X \left(X'X \right)^{-1} X' Z^{\perp X} \left[  (Z^{\perp X})' Z^{\perp X} \right] ^{-1} (Z^{\perp X})' = 0,
			\] 
			where I have used $X' Z^{\perp X} = 0$ (i.e. residuals are orthogonal to the regressors). 
			
			I pre-multiply (\ref{eq:decomp1}) by $Y'$, then post-multiply the result by $Y$, and finally divide through by $Y' M^0 Y$  to get
			\[
			\frac{Y'P_W M^0 P_W Y}{Y' M^0 Y} = \frac{Y' P_X M^0 P_X Y}{Y' M^0 Y} + \frac{Y' P_{Z^{\perp X}} M^0 P_{Z^{\perp X}} Y}{Y' M^0 Y}.
			\]
			Using (\ref{r2:yxz}) and (\ref{r2:yx}), we get
			\[
			R^2_{Y \sim X+Z} - R^2_{Y \sim X} = \frac{Y' P_{Z^{\perp X}} M^0 P_{Z^{\perp X}} Y}{Y' M^0 Y}.
			\]
			Since the right hand side is $R^2_{Y \sim Z^{\perp X}}$, we get (\ref{r2decomp-1}).
			
			We proceed by subtracting $R^2_{Y \sim Z}$ from both sides of the above equality. Using (\ref{r2:yz}), we get:
			\begin{equation}\label{eq:r2-diff}
				R^2_{Y \sim X+Z} - R^2_{Y \sim X} - R^2_{Y \sim Z}= \frac{Y' P_{Z^{\perp X}} M^0 P_{Z^{\perp X}} Y - Y' P_{Z} M^0 P_{Z} Y }{Y' M^0 Y} 
			\end{equation}
			Now, using the definition of $\eta_{X,Y,Z}$ in (\ref{def:etaxyz}), we see that the RHS of (\ref{eq:r2-diff}) is $\eta_{X,Y,Z}$. Hence, we get (\ref{r2decomp-2}):
			\[
			R^2_{Y \sim X+Z} - R^2_{Y \sim X} - R^2_{Y \sim Z}= \eta_{X,Y,Z}.
			\]
			
		\end{proof}
		
		\subsection{Proof of Theorem~\ref{thm:r2dzx}}
		\begin{proof}
			We know $\left| R_{D \sim Z|X_{-j}}\right|  = \sqrt{k_D} \left| R_{D \sim X_j|X_{-j}}\right| $. Now using the recursive definition of partial correlations \citep[equation 16, page~50]{cinelli-hazlett-2020}, we have
			\begin{align*}
				\left| R_{D \sim Z|X}\right| & = \frac{\left| R_{D \sim Z|X_{-j}} - R_{D \sim X_j|X_{-j}} R_{Z \sim X_j|X_{-j}}\right| }{\sqrt{1-R^2_{D \sim X_j|X_{-j}}} \sqrt{1-R^2_{Z \sim X_j|X_{-j}}}} \\
				& \geq \frac{\left| R_{D \sim Z|X_{-j}}\right|  - \left| R_{D \sim X_j|X_{-j}} R_{Z \sim X_j|X_{-j}}\right| }{\sqrt{1-R^2_{D \sim X_j|X_{-j}}} \sqrt{1-R^2_{Z \sim X_j|X_{-j}}}} \\
				& = \frac{\sqrt{k_D} \left| R_{D \sim X_j|X_{-j}}\right|  - \left| R_{D \sim X_j|X_{-j}} \right| \left| R_{Z \sim X_j|X_{-j}}\right| }{\sqrt{1-R^2_{D \sim X_j|X_{-j}}} \sqrt{1-R^2_{Z \sim X_j|X_{-j}}}} \\
				& = \frac{ \left| R_{D \sim X_j|X_{-j}}\right|  \left( \sqrt{k_D}- \left| R_{Z \sim X_j|X_{-j}}\right| \right) }{\sqrt{1-R^2_{D \sim X_j|X_{-j}}} \sqrt{1-R^2_{Z \sim X_j|X_{-j}}}}
			\end{align*} 
			where the second step uses the well-known result for real numbers: $|a-b| \geq |a| - |b|$. Hence, taking the square of both sides of the above inequality, using the definition of $k_D$ in (\ref{kd-partial-single}), and noting that 
			\[
			f^2_{D \sim X_j|X_{-j}} = \frac{R^2_{D \sim X_j|X_{-j}}}{1-R^2_{D \sim X_j|X_{-j}}}
			\]
			we get
			\begin{equation}
				R^2_{D \sim Z|X} \geq \alpha_D f^2_{D \sim X_j|X_{-j}}
			\end{equation}
			where
			\[
			\alpha_D = \frac{\left( \sqrt{k_D} - \left|  R_{Z \sim X_j | X_{-j}}\right|  \right)^2 }{1-R^2_{Z \sim X_j | X_{-j}}} \geq 0.
			\]
			A similar argument shows that 
			\begin{equation}
				R^2_{Y \sim Z|X} \geq \alpha_Y f^2_{Y \sim X_j|X_{-j}}
			\end{equation}
			where
			\[
			\alpha_Y = \frac{\left( \sqrt{k_Y} - \left|  R_{Z \sim X_j | X_{-j}}\right|  \right)^2 }{1-R^2_{Z \sim X_j | X_{-j}}} \geq 0.
			\]
		\end{proof}
		
	\subsection{Proof of Proposition~\ref{lem2}}
	\begin{proof}
		
		For any two real numbers, $x,y$, $\sign\left( x\right) = \sign\left( y \right) $ is equivalent to $|x+y|=|x|+|y|$. To see this, note $\left( x+y\right)^2 - \left( |x|+|y|\right)^2= 2\left( xy - |xy|\right) $. Thus, $\left( x+y\right)^2 = \left( |x|+|y|\right)^2$ iff $xy = |xy|$ iff $xy>0$ iff $\sign\left( x\right) = \sign\left( y \right) $. Now let $x=a-bc$ and $y=bc$ to get the result: $\sign\left( a-bc\right) = \sign\left( bc\right) $ iff
		 $|a-bc|=|a| - |bc|$.
	\end{proof}	
	
	\subsection{Proof of claims of section~\ref{sec:confd-increase}}\label{sec:app-confd-increase}
	
	Case~1: Sign of estimated and true treatment effect are same.\\
	
	\noindent
	In this case, we have
	\begin{equation}
		\frac{\widehat{\text{bias}}}{\hat{\tau}_{\text{res}}} <0.
	\end{equation} 
	To see this consider two possibilities: $\hat{\tau}>0$ or $\hat{\tau} < 0$. If $\hat{\tau}>0$, then $\hat{\tau}_{\text{res}} > \hat{\tau}>0 $. Hence, using ($\ref{def-bias}$), we see that $\widehat{\text{bias}}>0$. Since the signs of the estimated and true treatment effect are same, $\hat{\tau}_{\text{res}}>0$. Hence, $\frac{\widehat{\text{bias}}}{\hat{\tau}_{\text{res}}} >0$. On the other hand, if $\hat{\tau}\leq 0$, then $\hat{\tau}_{\text{res}} < \hat{\tau}<0 $, so that $\widehat{\text{bias}}<0$. Since the signs of the estimated and true treatment effect are same, $\hat{\tau}_{\text{res}}<0$. Hence, $\frac{\widehat{\text{bias}}}{\hat{\tau}_{\text{res}}} >0$. \\
	
	\noindent
	Case~2: Sign of estimated and true treatment effect are opposite.\\
	
	\noindent
	In this case, we have
	\begin{equation}
		\frac{\widehat{\text{bias}}}{\hat{\tau}_{\text{res}}} >0.
	\end{equation} 
	To see this consider, once again, two possibilities: $\hat{\tau}>0$ or $\hat{\tau} < 0$. If $\hat{\tau}>0$, then $\hat{\tau}_{\text{res}} < -\hat{\tau}<0 < \hat{\tau}$. Hence, $\widehat{\text{bias}}<0$, so that $\frac{\widehat{\text{bias}}}{\hat{\tau}_{\text{res}}} >0$. On the other hand, if $\hat{\tau} < 0$, then $\hat{\tau}_{\text{res}} > -\hat{\tau} >0 > \hat{\tau} $, so that $\widehat{\text{bias}}>0$. Hence, $\frac{\widehat{\text{bias}}}{\hat{\tau}_{\text{res}}} >0$.

	\end{appendices}
	
\end{document}